\documentclass[useAMS,twocolumn,trackchanges]{aastex63}

\usepackage{amsmath,amssymb}
\usepackage{natbib}
\usepackage{epsfig}
\usepackage{times}
\usepackage{rotate}
\usepackage{hyperref}
\usepackage{color,url}
\usepackage{subfigure}
\usepackage{cprotect}
\usepackage{xspace}
\usepackage{enumerate}
\usepackage{fancyhdr}
\usepackage{appendix}
\usepackage[pass]{geometry}

\newcommand{\simgt}{\lower.5ex\hbox{$\; \buildrel > \over \sim \;$}}
\newcommand{\simlt}{\lower.5ex\hbox{$\; \buildrel < \over \sim \;$}}

\newcommand{\myemail}{mskobayashi@astr.tohoku.ac.jp, masato.kobayashi@nagoya-u.jp}

\newcommand{\m}{${\rm M}_{\rm 200b}$ }

\newcommand{\eg}{e.g.\xspace}
\newcommand{\cf}{c.f.\xspace}
\newcommand{\ie}{i.e.\xspace}
\newcommand{\etc}{etc.}

\newcommand{\msun}{\mbox{${\rm M_{\odot}}$}\xspace}

\newcommand{\kms}{{\rm km \, s^{-1}}}

\newcommand{\cmkk}{\mbox{${\rm cm^{-3}}$}}

\newcommand{\myxshock}{\langle x_{\rm  shock} \rangle}
\newcommand{\myvin}{V_{\rm in}}
\newcommand{\myvinp}{V_{\rm in,pre}}
\newcommand{\myvshock}{V_{\rm shock}}

\newcommand{\myvej}{v_{\rm ej}}

\newcommand{\mytsgl}{t_{\rm sg}}
\newcommand{\mytstop}{t_{\rm stop}}
\newcommand{\MachN}{\mathcal{M}}
\newcommand{\MachNs}{\mathcal{M}_{\rm shock}}
\newcommand{\myCooL}{\mathcal{L}}
\newcommand{\mygamma}{\gamma_{\rm eff}}
\newcommand{\gor}{\gamma_0}

\newcommand{\dvws}{\langle \delta v_{\rm dw}^2 \rangle}
\newcommand{\dvw}{\sqrt{\dvws}}
\newcommand{\ntanh}{n_{\rm smx}}

\newcommand{\myrj}{r_{\rm d}}
\newcommand{\mymgas}{m_{\rm gas}}
\newcommand{\myboltz}{k_{\rm B}^{}}
\newcommand{\mycs}{C_{\rm s}^{}}
\newcommand{\mycswnm}{C_{\rm s, WNM}^{}}

\newcommand{\mytaucool}{\tau_{\rm cool}^{}}
\newcommand{\mylcool}{\lambda_{\rm cool}^{}}
\newcommand{\mydrhop}{\Delta \rho_{0}}
\newcommand{\myaa}{\alpha_1}
\newcommand{\myab}{\alpha_2}
\newcommand{\myac}{\alpha_3}

\received{--}
\revised{--}
\accepted{--}
\submitjournal{ApJ}

\begin{document}

%-------- TITLE  ---------------------

    \title{BIMODAL BEHAVIOR AND CONVERGENCE REQUIREMENT IN MACROSCOPIC PROPERTIES \\
           OF THE MULTIPHASE INTERSTELLAR MEDIUM FORMED BY ATOMIC CONVERGING FLOWS}
    \shorttitle{Bimodal properties of the ISM}
    \shortauthors{Masato I.N. Kobayashi}


%-------- AUTHORS  ---------------------


    \correspondingauthor{Masato I.N. Kobayashi}
    \email{\myemail}
    \author[0000-0003-3990-1204]{Masato I.N. Kobayashi}
    \affiliation{Astronomical Institute, Graduate School of Science, Tohoku University, Aoba, Sendai, Miyagi 980-8578, Japan}\\
    %\affiliation{Department of Earth and Space Science, Graduate School of Science, Osaka University, 1-1, Machikaneyama-cho, Toyonaka, Osaka 560-0043, Japan}\\
    \author{Tsuyoshi Inoue}
    \affiliation{Division of Particle and Astrophysical Science, Graduate School of Science, Nagoya University, Aichi 464-8602, Japan}
    \author{Shu-ichiro Inutsuka}
    \affiliation{Division of Particle and Astrophysical Science, Graduate School of Science, Nagoya University, Aichi 464-8602, Japan}
    \author[0000-0001-8105-8113]{Kengo Tomida}
    \affiliation{Astronomical Institute, Graduate School of Science, Tohoku University, Aoba, Sendai, Miyagi 980-8578, Japan}\\
    %\affiliation{Department of Astrophysical Sciences, Princeton University, Princeton, NJ 08544, USA}\\
    \author[0000-0002-2707-7548]{Kazunari Iwasaki}
    \affiliation{Center for Computational Astrophysics, National Astronomical Observatory of Japan, Mitaka, Tokyo 181-8588, Japan}\\
    \author[0000-0002-6907-0926]{Kei E. I. Tanaka}
    \affiliation{ALMA Project, National Astronomical Observatory of Japan, Mitaka, Tokyo 181-8588, Japan}\\
    \shortauthors{M. I. N. Kobayashi et al.}
   
    %\maketitle
    %\setcounter{footnote}{6}
    %\label{firstpage}

%-------- ABSTRACT  ---------------------

\begin{abstract}
We systematically perform hydrodynamics simulations of 20 $\kms$ converging flows of the warm neutral medium (WNM)
to calculate the formation of the cold neutral medium (CNM),
especially focusing on 
the mean properties of the multiphase interstellar medium (ISM),
such as 
the average shock front position and
the mean density on a 10 pc scale.
Our results show that the convergence in those mean properties requires
0.02 pc spatial resolution
that 
resolves 
the cooling length of the thermally unstable neutral medium (UNM)
to follow the dynamical condensation from the WNM to CNM.
%fully resolves
%the typical cooling length 
%on which the phase transition occurs from the WNM to CNM\@.
We also find that two distinct post-shock states
appear in the mean properties
depending on the amplitude of the upstream WNM density fluctuation
$\mydrhop$ ($=\sqrt{\langle \delta \rho_0^2 \rangle}/\rho_0$).
When $\mydrhop > 10$ \%, 
the interaction between shocks and density inhomogeneity
leads to a strong driving of the post-shock turbulence
of $> 3$ km s$^{-1}$, which
dominates the energy budget in the shock-compressed layer.
The turbulence prevents 
the dynamical condensation by cooling 
and the following CNM formation,
and 
the CNM mass fraction remains as $\sim 45$ \%.
%in 3 Myr.
In contrast, when $\mydrhop \leq 10$ \%, 
the shock fronts maintain an almost straight geometry 
and CNM formation efficiently proceeds, resulting in 
the CNM mass fraction of $\sim 70$ \%.
%in 3 Myr.
The velocity dispersion is limited to 
the thermal-instability mediated level of $\sim 2$ -- $3$ km s$^{-1}$
and the layer is supported by both turbulent and thermal energy equally.
%We propose that 
%the multiphase ISM formed by the WNM converging flow
%can be modeled as a one-phase ISM in a form of effective equation of state $P\propto \rho^{\mygamma}$,
%where $\mygamma$ varies from $0.9$ (for large pre-shock $\mydrhop$) to $0.7$ (for small pre-shock $\mydrhop$)
We also propose an effective equation of state that models the multiphase ISM formed by the WNM converging flow as a one-phase ISM.

\end{abstract}

%\keywords{ISM: Interstellar phases: WNM, ISM: Interstellar phases: CNM}

%\maketitle
%\setcounter{footnote}{6}

%-------- KEY WORDS  ---------------------



\section{Introduction}
\label{sec:intro}
Molecular clouds are formation sites of stars
\citep{Kennicutt2012}
and the formation process
of molecular clouds is essential to understand
the initial condition of star formation.
Since the galactic disks are largely
occupied by the warm neutral medium (WNM: $\simeq$ 6000 K and $\simeq$ 1 $\cmkk$),
the phase transition from the WNM to the cold neutral medium (CNM: $\simeq$ 100 K and $\simeq$ 100 $\cmkk$)
is the initial step in the formation of molecular clouds ($\simeq$ 10 K and $\geq$ 100 $\cmkk$).
The thermal instability due to radiative cooling and heating
significantly influences this phase transition dynamics
\citep{Field1965,Field1969,Zeldovich1969,Wolfire1995,Wolfire2003},
and is believed to be triggered by supersonic flows
originated in supernovae \citep{McKee1977,Chevalier1977,Chevalier1999},
superbubbles \citep{McCray1979,Tomisaka1981,Tomisaka1986b,Kim2017b,Ntormousi2017},
galactic spirals \citep{Shu1972,Wada2011,Baba2017,KimWT2020},
and galaxy mergers \citep{Heitsch2006b,Arata2018}.

Many authors have studied the dynamical condensation 
through the thermal instability in a shock-compressed layer 
formed by supersonic converging flows, whose importance is
initially highlighted with one-dimensional simulations 
\citep[\eg,][]{Hennebelle1999,Koyama2000}.
This converging-flow configuration is a technical analogue 
to easily calculate the thermal instability
in the post-shock rest frame,
instead of tracking the post-shock interstellar medium (ISM) propagating with a shock front in space
\citep[see][]{Koyama2000}.
Multi-dimensional simulations are later performed
in two dimensions \citep{Koyama2002,Audit2005,Heitsch2005,Hennebelle2007a}
and in three dimensions
\citep{Heitsch2006b,vazquezsemadeni2006,vazquezsemadeni2007,Audit2008}.
They essentially form the multiphase ISM with supersonic turbulence 
consistent with observations \citep[\eg,][]{Larson1981,Heyer2004}.
There have been also converging-flow studies that extensively investigate the effect of magnetic fields 
(\citealt{Hennebelle2000,Inoue2008,Inoue2009,Hennebelle2008,Heitsch2009,vazquezsemadeni2011,Inoue2012,Valdivia2016,Inoue2016,Iwasaki2018}; 
\cf, \citealt{vanLoo2007,vanLoo2010})
and gas-phase metallicity \citep{Inoue2015},
reporting that the flow direction is redirected by magnetic fields and
the critical metallicity is $\sim 0.04 Z_{\odot}$\footnote[6]{$Z_{\odot}$
represents the solar metallicity.}, above which the ISM becomes biphasic. 

Theoretical studies
of the multiphase ISM obtained
in those converging-flow simulations
are 
promising to fill the spatial and timescale gap in numerical studies 
between the galactic-disk evolution on $\geq 1$ kpc scales over $100$ Myr
and the formation of individual molecular cloud cores/stars 
on $\leq 0.1$ pc scales over a few Myr.
For example, numerical simulations of the entire galactic disks
start to achieve the mass (spatial) resolution down to 
$10^4\,\msun$ ($< 10$ pc) \citep{Wada1999,Baba2017},
but need an aid of some zooming techniques
to simultaneously resolve individual cloud cores.
Also simulations of a fraction of galactic disks
are performed on $\sim$ kpc$^3$ volume
over a few 100 Myr to investigate the ISM evolution driven by
multiple supernovae
\citep[\eg,][]{Gent2013,Hennebelle2014,Walch2015,Girichidis2016,Gatto2017,Kim2017,Colling2018,
KimWT2020}, whose spatial resolutions are typically a few to $10$ pc
(see also \citealt{Bonnell2013,Hennebelle2018} for zoom-in simulations).
Therefore converging-flow simulations on $10$ pc scales 
can be utilized
to provide sub-grid models
for large-scale simulations
to consistently calculate time-evolution of the multiphase ISM
below the spatial resolution (\eg, 
time-evolution of the mean density averaged on $10$ pc scales).

However, in the converging-flow simulations, authors employ different implementations of 
perturbation that initiates the thermal instability and different perturbation amplitudes.
For example, 
a fluctuation is introduced
in the upstream flow 
density \citep[\eg,][]{Koyama2002,Inoue2012,CarrollNellenback2014},
velocity \citep[\eg,][]{Audit2005,vazquezsemadeni2006},
or a sinusoidal interface where
two flows initially collide \citep[\eg,][]{Heitsch2005}.
Therefore, it is not clear yet whether 
detailed initial settings of the flow 
impact not only the detailed
ISM properties
but also the corresponding mean properties 
which large-scale simulations need in their sub-grid model.
In addition, previous discussions on the spatial resolution of converging-flows
put emphasis on whether they resolve dense CNM structures and turbulent structures
\citep[\eg,][]{Koyama2004,Audit2005,Inoue2015}
and discussions on the mean properties are still limited; \eg, 
column density \citep{vazquezsemadeni2006} and chemical abundance \citep{Joshi2019}.
Therefore, it is still a remaining task
for converging-flow simulations
to systematically investigate 
the spatial resolution 
required for the convergence of the mean properties,
and to reveal how detailed conditions of the flow
impact such mean properties,
under a fixed method of perturbation seeding.

In this article, 
we perform converging flow simulations
on a 10 pc scale over 3 Myr
with the spatial resolution up to $0.01$ pc.
We employ
an upstream density inhomogeneity $\Delta \rho_0$ as a perturbation seed.
This is motivated by the fact that
density inhomogeneity exists
on all spatial scales in the diffuse ISM
\citep{Armstrong1995}
down to star-forming regions
\citep{Schneider2013,Schneider2016},
where supersonic flows of supernova remnants and H {\sc ii} regions are naturally expected to 
expand through such density fluctuation \citep{Inoue2012c,Kim2015}.
We systematically vary the amplitude of the upstream density fluctuation 
$\mydrhop$
and the spatial resolution $\Delta x$ to 
reveal the dependence of the physical properties of the multiphase ISM
on those conditions.
We also change the perturbation phase 
under a fixed 
perturbation power spectrum
and spatial resolution,
aiming at revealing the statistical variation
that the mean properties intrinsically have
but large-scale simulations are not able to directly calculate.

The rest of this article is organized as follows.
In Section~\ref{sec:method}, we describe the method 
to perform our simulations and list the parameter space
that we study.
In Section~\ref{sec:results}, we show the results of 
our simulations, first by exploring the convergence with $\Delta x$,
and variety depending on $\mydrhop$.
We discuss our results and provide additional analyses
in Section~\ref{sec:discussions}, and summarize this article in Section~\ref{sec:concl}.

\section{Method}
\label{sec:method}

\subsection{Basic Equations}
\label{subsec:code}
\begin{figure}\centering{
\includegraphics[width=0.70\columnwidth,keepaspectratio]{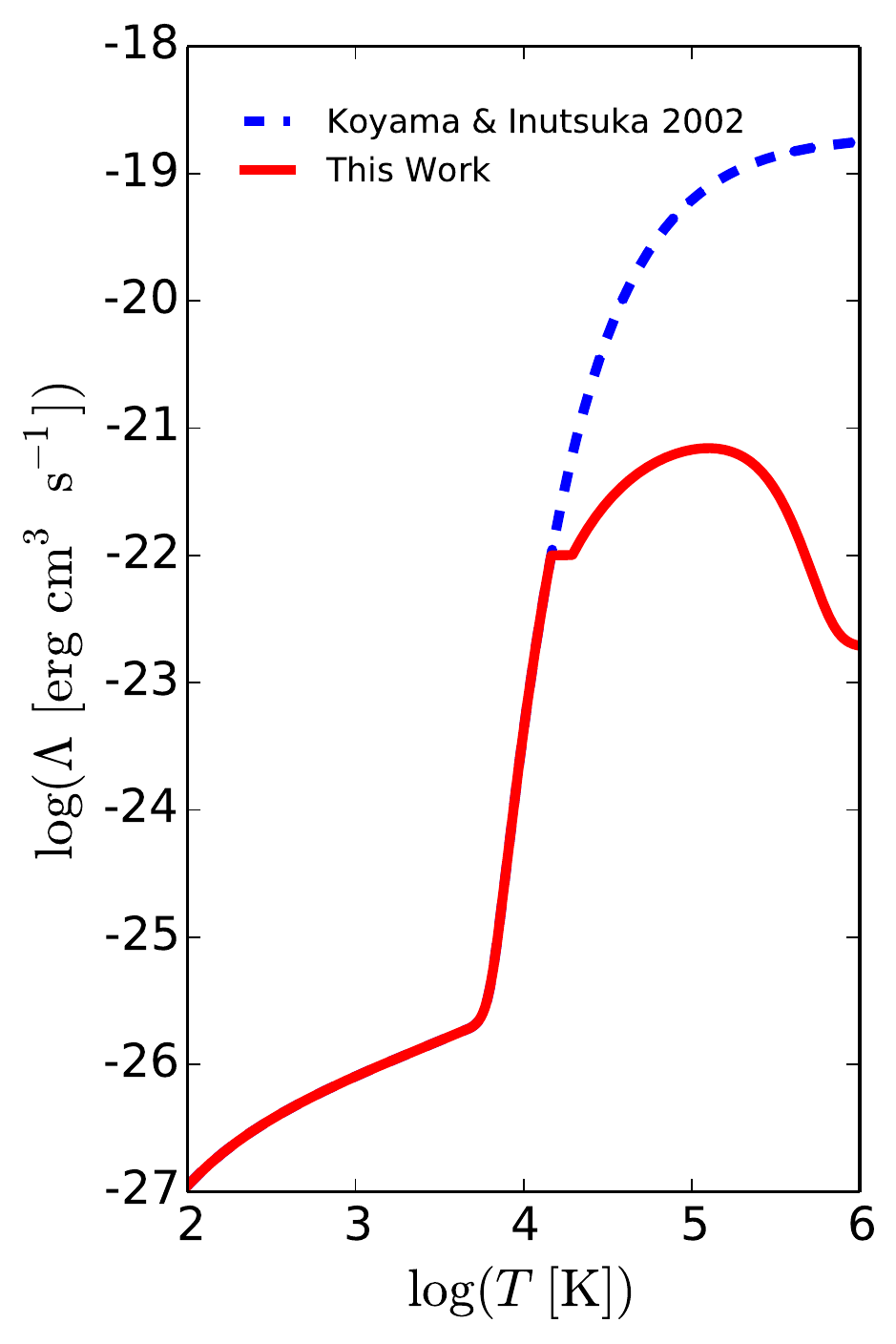}}
\caption{Comparison between the cooling functions from 
    \cite{Koyama2002} (blue dashed) 
    and from this work (red solid). 
    Our modifications on \cite{Koyama2002} are motivated 
    to calculate the cooling rate in the high-temperature regime,
    especially reflecting the facts that the thermal ionization 
    of neutral hydrogen ($T \gtrsim 14,577$ K) 
    and line emission by ionized carbon and oxygen (peaks at $T\sim 10^5$ K)
    \citep{Cox1969,Dalgarno1972}.}
\label{fig:f1}
\end{figure}

We utilize the hydrodynamics part from the magneto-hydrodynamics code by 
\cite{Inoue2008}. This solves equations of hydrodynamics  
based on the second-order Godunov scheme \citep{vanLeer1979}. Heating and cooling 
are explicitly time-integrated and have the second order accuracy. %in both space and time.
We solve the following basic equations: 
\begin{eqnarray}
    &&\frac{\partial \rho}{\partial t} + \nabla_{\mu}^{} (\rho v_{\mu}) = 0 \,, \label{eq:CoM}\\
    &&\frac{\partial (\rho v_{\mu})}{\partial t} 
       + \nabla_{\nu} (P\delta_{\mu\nu} + \rho v_{\nu}v_{\mu}) = 0 \,, \label{eq:EoM}\\
    &&\frac{\partial e}{\partial t} + \nabla_{\mu} \left( (e + P)v_{\mu} \right) 
       = \nabla_{\mu}(\kappa(T)\nabla_{\mu} T) -\rho \myCooL \label{eq:EE}\,.
\end{eqnarray}
Here, $\rho$ represents the mass density, $v$ represents the velocity,
$P$ is the thermal pressure, $T$ is the temperature, and 
$\nabla_{\mu} = \partial/\partial \mu$,
where $\mu$ spans $x, y,$ and $z$.
The total energy density,
$e$, is given as $e = P/(\gamma-1) + \rho v^2/2$
where $\gamma$ means the ratio of the specific heat ($=5/3$). We implement the thermal conductivity, $\kappa$, 
as $\kappa(T) = 2.5 \times 10^3 \, T^{0.5} \, 
\mathrm{erg \, cm^{-1} \, s^{-1} \, K^{-1}}$
by considering collisions
between hydrogen atoms \citep{Parker1953}.
The net cooling rate per mass $\myCooL$ is the integration of 
heating and cooling processes (see Figure~\ref{fig:f1}).
We utilize the following net cooling function,
\begin{equation}
\begin{aligned}
    &\rho \myCooL = -\left(\frac{\rho}{\mymgas}\right) \Gamma +
       \left(\frac{\rho}{\mymgas}\right)^2 \Lambda \,, \\
    &\Gamma = 2 \times 10^{-26} \, \mathrm{erg \, s^{-1}} \,, \\
    &\frac{\Lambda(T)}{\Gamma} = \\
    &\begin{cases}
        10^7 \exp\left( \frac{-118400}{T+1000} \right) \\
        \hspace{0.3cm} + 1.4\times 10^{-2} \sqrt{T} \exp \left( \frac{-92}{T} \right) \, \mathrm{cm^3} 
        \, (\mathrm{for} \, T\leq14,577 \,\mathrm{K})\,, \\ %\label{eq:KI2002_cool}\,, \\
        5 \times 10^3 + 1.4\times 10^{-2} \sqrt{T} \exp \left( \frac{-92}{T} \right) \, \mathrm{cm^3} \\
        \hspace{2.1cm}  (\mathrm{for} \, 14,577\,\mathrm{K}< T \leq 19,449 \,\mathrm{K})\,, \\ %\label{eq:KI2002_cool}\,, \\
        3.75 \times 10^4 \left( 1 -\tanh \left(\frac{T-2\times10^5}{2\times10^5}\right) \right) \exp \left(\frac{-5\times10^4}{T}\right) \\
        \hspace{0.3cm} +10^3 \exp \left( \frac{-5\times10^4}{T}\right) \, \mathrm{cm^3} 
        \, (\mathrm{for} \, T > 19,449 \,\mathrm{K})\,, \\ %\label{eq:cool_modified}\,, \\
    \end{cases}
    \label{eq:hcfunc}
\end{aligned}
\end{equation}
where $\Gamma$ is the heating rate,
$m_{\rm gas}$ is the mean particle mass,
and all the $T$ represents the temperature in the unit of Kelvin.
The heating process, $\Gamma$, 
comes from photoelectric heating by polycyclic aromatic hydrocarbons.
The cooling process, $\Lambda(T)$, is proposed in 
\cite{Koyama2002},
which is based on detailed calculation 
of heating and cooling rates in optically thin ISM from 
\cite{Koyama2000}.
We use this formula in $T \leq 14,577$ K,
which mainly consists of two terms;
the Ly$\alpha$ cooling (the first term) and the C$_{\rm II}$ cooling (the second term).
We modify this function at higher temperature regime
as shown equations for $T > 14,577$ K,
by considering
He, C, O, N, Ne, Si, Fe, and Mg (\citealt[][]{Cox1969,Dalgarno1972};
Figure~\ref{fig:f1})\footnote[7]{We refer the readers to
see \cite{Micic2013} that the detail choice of cooling function 
does not alter the cold gas mass $T<300$ K,
where the author employs the spatial resolution of $0.03$ pc,
close to our current simulations.}.
The typical cooling timescale of the injected WNM, $\mytaucool$,
is $1.3$ Myr under this cooling function.
This timescale becomes shorter once the injected WNM is compressed
by the shock\footnote[8]{We refer the readers to see 
other references (\eg, \citealt{Inoue2012}) for
more detailed chemical networks of the ISM during molecular cloud formation phase.}.

We define the thermally unstable neutral medium (UNM) as 
$(\partial (\mathcal{L}/T)/\partial T)_P <0$ \citep[\eg,][]{Balbus1986,Balbus1995},
and the gas state warmer (colder) than the UNM is classified as the WNM (CNM).
We will use this definition hereafter when we analyze the mass fraction of each phase.
Although the boundary between WNM, UNM, and CNM depends on both  the density and temperature,
the three phases roughly correspond to 
the WNM as $T\geq5000$ K, the UNM as $100 \leq T <5000$ K, and
the CNM as $T<100$ K based on Equation~\ref{eq:hcfunc}.
%The boundary between WNM, UNM, and CNM depends both on the density and temperature,
%but the three phases roughly correspond to 
%the WNM as $T\geq5000$ K, the UNM as $100 \leq T <5000$ K, and 
%the CNM as $T<100$ K. We will use this definition hereafter.
%Therefore, based on Equation~\ref{eq:hcfunc}, let us define the WNM, UNM, and CNM as follows:
%the WNM as $T\geq5000$ K, the UNM as $100 \leq T <5000$ K, and 
%the CNM as $T<100$ K. We will use this definition hereafter.}

\begin{figure*}\centering{
    \includegraphics[scale=0.6,keepaspectratio]{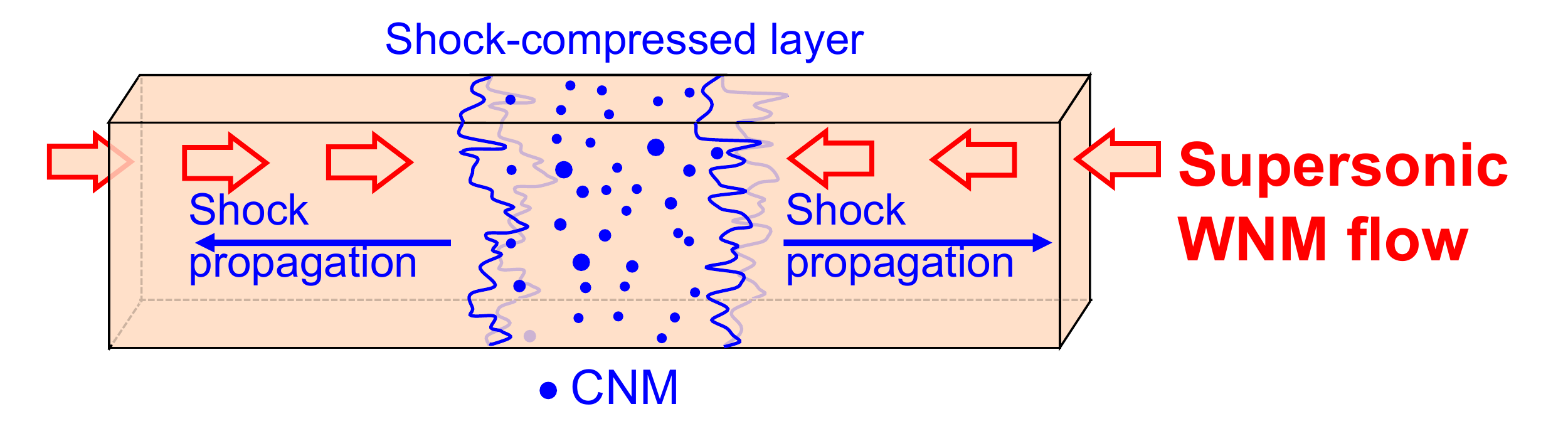}}
    \caption{Schematic figure of our converging flow simulations.
    Supersonic WNM flows are continuously injected at the $x$ boundaries
    and flow inwards (red arrows).
    The shock-compressed layer forms at the center of the simulation box,
    sandwiched by two shock fronts (blue solid lines).
    It becomes thicken while the flow continues
    (blue arrows). Dense clumps whose density correspond to CNM
    form in the layer (blue points).}
\label{fig:f2}
\end{figure*}

\subsection{Setups and Shock Capturing}
\label{subsec:setup}
\begin{table*}
    \caption{Studied Parameters}
    \centering{
        \begin{tabular}{c|cccccc}
            \hline
            \hline
            Runs & \multicolumn{6}{c}{Parameters}\\
\hline
& Adiabatic (A)          & Resolution    & $\Delta \rho_0$                                     & Phase     & $\gamma$ & Figures\\
& or Heating             & $\Delta  x$   & $=\sqrt{\langle \delta \rho_0^2 \rangle} / \rho_0$  & $\alpha$  &        & \\
&    \& Cooling (C)      & $($ pc $)$    & $($ \% $)$                                          &          &        & \\
& or Effective index (E) &               &                                                     &           &        & \\
 \hline 
 \hline
A2D0000  & A   & 0.02   & 0     & N/A                        & $\gor=$ 1.67   & ~\ref{fig:f9} \\ 
C2D0000  & C   & 0.02   & 0     & N/A                        & $\gor=$ 1.67   & ~\ref{fig:f9}, and ~\ref{fig:f3f4f5f6f7f8} \\ 
C2D1000  & C   & 0.02   & 100   & $\myaa$, $\myab$, $\myac$  & $\gor=$ 1.67   & ~\ref{fig:f3f4f5f6f7f8}, ~\ref{fig:f18f19f20f21}, ~\ref{fig:f22f23}, ~\ref{fig:f24f25f26f27}, ~\ref{fig:f28}, ~\ref{fig:f29f30}, ~\ref{fig:f32f33}\\ 
C8D0316  & C   & 0.08   & 31.6  & $\myaa$, $\myab$, $\myac$  & $\gor=$ 1.67   & ~\ref{fig:f10f11f12f13}, ~\ref{fig:f14f15f16f17}, ~\ref{fig:f18f19f20f21} \\ 
C4D0316  & C   & 0.04   & 31.6  & $\myaa$, $\myab$, $\myac$  & $\gor=$ 1.67   & ~\ref{fig:f10f11f12f13}, ~\ref{fig:f14f15f16f17}, ~\ref{fig:f18f19f20f21} \\ 
C2D0316  & C   & 0.02   & 31.6  & $\myaa$, $\myab$, $\myac$  & $\gor=$ 1.67   & ~\ref{fig:f3f4f5f6f7f8}, ~\ref{fig:f9}, ~\ref{fig:f10f11f12f13}, ~\ref{fig:f14f15f16f17}, ~\ref{fig:f18f19f20f21}, ~\ref{fig:f22f23}, ~\ref{fig:f28}, ~\ref{fig:f29f30}, ~\ref{fig:f32f33} \\ 
C1D0316  & C   & 0.01   & 31.6  & $\myaa$                    & $\gor=$ 1.67   & ~\ref{fig:f10f11f12f13}, ~\ref{fig:f14f15f16f17},~\ref{fig:f18f19f20f21} \\ 
C8D0100  & C   & 0.08   & 10    & $\myaa$, $\myab$, $\myac$  & $\gor=$ 1.67   & ~\ref{fig:f18f19f20f21}\\ 
C4D0100  & C   & 0.04   & 10    & $\myaa$, $\myab$, $\myac$  & $\gor=$ 1.67   & ~\ref{fig:f18f19f20f21}\\ 
C2D0100  & C   & 0.02   & 10    & $\myaa$, $\myab$, $\myac$  & $\gor=$ 1.67   & ~\ref{fig:f3f4f5f6f7f8}, ~\ref{fig:f18f19f20f21}, ~\ref{fig:f22f23}, ~\ref{fig:f28}, ~\ref{fig:f29f30}, ~\ref{fig:f32f33} \\ 
C1D0100  & C   & 0.01   & 10    & $\myaa$                    & $\gor=$ 1.67   & ~\ref{fig:f18f19f20f21}\\ 
C8D0031  & C   & 0.08   & 3.16  & $\myaa$, $\myab$, $\myac$  & $\gor=$ 1.67   & ~\ref{fig:f10f11f12f13}, ~\ref{fig:f14f15f16f17}, ~\ref{fig:f18f19f20f21} \\ 
C4D0031  & C   & 0.04   & 3.16  & $\myaa$, $\myab$, $\myac$  & $\gor=$ 1.67   & ~\ref{fig:f10f11f12f13}, ~\ref{fig:f14f15f16f17}, ~\ref{fig:f18f19f20f21} \\ 
C2D0031  & C   & 0.02   & 3.16  & $\myaa$, $\myab$, $\myac$  & $\gor=$ 1.67   & ~\ref{fig:f3f4f5f6f7f8}, ~\ref{fig:f10f11f12f13}, ~\ref{fig:f14f15f16f17}, ~\ref{fig:f18f19f20f21}, ~\ref{fig:f22f23}, ~\ref{fig:f29f30}  \\ 
C1D0031  & C   & 0.01   & 3.16  & $\myaa$                    & $\gor=$ 1.67   & ~\ref{fig:f10f11f12f13}, ~\ref{fig:f14f15f16f17}, ~\ref{fig:f18f19f20f21} \\ 
C2D0010  & C   & 0.02   & 1     & $\myaa$, $\myab$, $\myac$  & $\gor=$ 1.67   & ~\ref{fig:f3f4f5f6f7f8}, ~\ref{fig:f18f19f20f21}, ~\ref{fig:f22f23}, ~\ref{fig:f24f25f26f27}, ~\ref{fig:f28}, ~\ref{fig:f29f30}, ~\ref{fig:f32f33} \\ 
C2D0003  & C   & 0.02   & 0.316 & $\myaa$, $\myab$, $\myac$  & $\gor=$ 1.67   & ~\ref{fig:f3f4f5f6f7f8}, ~\ref{fig:f22f23}, ~\ref{fig:f29f30}  \\ 
C2D0001  & C   & 0.02   & 0.1   & $\myaa$, $\myab$, $\myac$  & $\gor=$ 1.67   & ~\ref{fig:f3f4f5f6f7f8}, ~\ref{fig:f18f19f20f21}, ~\ref{fig:f22f23}, ~\ref{fig:f29f30} \\ 
E2D0000  & E   & 0.02   & 0     & N/A                        & $\gor=$ 1.67   & \\
         &     &        &       &                            & \hspace{6mm} \& $\mygamma=0.863$ & ~\ref{fig:f32f33} \\ 

            \hline
            \hline
        \end{tabular}
    }\par
    \bigskip
            \textbf{Note.} 45 cases (out of total 69 cases) that we will present in this article.
              Details of individual parameters are described in Section~\ref{sec:method}. 
              The run names represent the following first three columns.
              Each column represents as follows; 
              The first column indicates whether we calculate adiabatic fluid (indicated as ``A''),
              or include heating and cooling processes based on Equation~\ref{eq:hcfunc} (indicated as ``C''),
              or using the effective index $\mygamma$ (indicated as ``E'').
              Resolution shows the spatial resolution.
              Phases $\myaa$, $\myab$, and $\myac$ represent three different random phases for initial density fluctuation
              ($\alpha_{k_y,k_z}$ in Equation~\ref{eq:rho_init}), 
              where N/A represents no fluctuation (\ie, $\sqrt{\langle \delta \rho_0^2 \rangle}=0.0$).
              $\gamma$ lists the original specific heat $\gor$ and the effective index $\mygamma$.
              Figures list the corresponding figure numbers.
    \label{table:cases}
\end{table*}

Figure~\ref{fig:f2} shows a schematic view of our simulations.
The shock-compressed layer forms at the box center
sandwiched by two shock fronts, and
becomes thicken while the flow continues as the two shocks propagate outwards.
Our three-dimensional simulation box has its size of $L_{x,y,z}=20,10,10$ pc,
which is a typical size of giant molecular clouds in the Milky Way galaxy.
Here $x$ is defined as the flow direction.
The left and right parts of the WNM have the velocity in the opposite direction
with $\left| v_x \right|= 20$ km s$^{-1}$, colliding at the box center.
The relative velocity between these two flows is thus $40\, \kms$ in this $20\, \kms$ + $20\,\kms$ collision.
This velocity is likely even faster in galaxy mergers (\eg, $>100\, \kms$),
but by employing 20 km s$^{-1}$, we opt to focus on more common situations in galactic disks 
(\eg, 
the late phase of supernova remnants expansions, H {\sc ii} region expansions,
and normal shock due to galactic spirals\footnote[9]{See also 
\cite{Ho2019} for recent \texttt{EAGLE} cosmological simulations,
which indicate that $20-60 \, \kms$ are the typical velocities 
with which cold gas $< 2.5 \times 10^5$ K accrete 
onto galaxies whose stellar mass is $\sim 10^{10} \, \msun$.}).
For the initial velocity field, we may simply flip the sign 
of the velocity at $x=L_x/2$.
However, as a more conservative approach
to avoid any artifacts that may arise from 
such a step-function collision,
we apply $\tanh$ smoothing as
\begin{equation}
    v_x = \myvin \tanh\left(\frac{x-L_x/2}{\ntanh \Delta x}\right) \,,
    \label{eq:tanh}
\end{equation}
so that the initial collision occurs smoothly. Here $\myvin = 20$ km s$^{-1}$ and
$\Delta x$ is the mesh size.
$\ntanh$ is chosen such that the physical scale of this smoothing is constant as
$\ntanh \Delta x = 0.78$ pc 
between different resolution runs.

The WNM is continuously injected through the two $x$ boundaries at $x=0$ and $20$ pc
until the calculation ends at 3 Myr,
whereas $y$ and $z$ boundaries have the periodic boundary condition.
The WNM flow is in a thermally stable phase
having the mean number density $n_0=0.57 \,\cmkk$
and pressure $P_0/\myboltz= 3500 \, \mathrm{K \, cm^{-3}}$.
The corresponding mean temperature, sound speed, and dynamical pressure are $6141$ K, $\mycs=8.16\,\kms$, and
$2.6 \times 10^4 \, \mathrm{K\,cm^{-3}}$, respectively,
where we use $\rho_0 = n_0 \mu_{\rm M}^{} m_{\rm p}$ with $\mu_{\rm M}^{} = 1.27$ as the mean molecular weight
\citep[\cf,][]{Inoue2008,Inoue2012}.

The mass density of the WNM flow has a fluctuation as $\rho (x,y,z) = \rho_0 +  \delta\rho_0(x,y,z)$:
\begin{eqnarray}
    \delta \rho_0(x,y,z) &=& \sum_{k_x,k_y,k_z} A(k) \nonumber  \\
     &&\times \sin (k_x^{}x+k_y^{}y+k_z^{}z+\alpha_{k_x^{},k_y^{},k_z^{}}^{}) \label{eq:rho_init} \,. 
\end{eqnarray}
$k_x^{}$, $k_y^{}$ and $k_z^{}$ are the wave numbers in $x$, $y$ and $z$-directions as
$k_x = 2\pi l_{x}/(L_{x}/2)$, $k_y = 2\pi l_{y}/L_{y}$ and $k_z = 2\pi l_{z}/L_{z}$ 
while the integers $l_{x}$, $l_{y}$ and $l_{z}$
span from $-32$ to $32$. 
$A(k)$ is set 
such that
the density power spectrum follows
the Kolmogorov spectrum $P_{\rho}(k)\propto k^{-11/3}$ 
where $k = \sqrt{k_x^2+k_y^2+k_z^2}$ \citep{Kolmogorov1941,Armstrong1995}.
As indicated in the definition of $k_x^{}$, 
we generate this fluctuation over a 10 pc $\times$ 10pc $\times$ 10pc volume
so that the initial left-half ($x=0$ -- $10$ pc) and right-half ($x=10$ -- $20$ pc) 
of the WNM have the same density distribution
(so as the temperature distribution does to achieve the initial pressure equilibrium).
The WNM flow injected from the boundaries also follow this density distribution
as the flows move inward, and the WNM accrete onto the shock-compressed layer always with this density fluctuation.

To identify the shock-compressed layer, 
we search cells along $x$ direction at every given $(y,z)$ inward from both $x$ boundaries, 
and label the first cells with $P>1.3P_0$.
We calculate the arithmetic mean of the distances from the box-center to these positions 
as the representative shock front position:
\begin{equation}
    \myxshock = \frac{\sum_{y,z}^{} \left( x_{\rm R}^{}(y,z)- x_{\rm L}^{}(y,z)\right)}{2N_{yz}}\,,
    \label{eq:xshock}
\end{equation}
where the subscripts $\mathrm{R}$ and $\mathrm{L}$ denote the two shock front positions 
(the right and left side of the layer), and $N_{yz}$ denotes the total number of cells on $y$-$z$ plane.
The factor of 2 in the denominator means that there are two shocks
and the overall mean position of the shock front is the half of $x_{\rm R} - x_{\rm L}$ at every given $(y,z)$.
We then measure the mean density within the shock-compressed layer $\langle n \rangle$
accordingly, as the gas density averaged over the entire volume of the shock-compressed layer
defined above.

\subsection{Range of Systematic Study}
\label{subsec:ss}
As a systematic study, we opt to vary two properties: 
the amplitude of the density fluctuation
$\langle A \rangle$ and the spatial resolution $\Delta x$.
We vary $\langle A \rangle$ such that 
the mean dispersion of the density $\mydrhop = \sqrt{\langle \delta \rho_0^2 \rangle }/ \rho_0$
spans as $100$, $31.6$, $10$, $3.16$, $1$, $0.361$, $0.1$ \%.
This $100$ \% is motivated by the fact that
density inhomogeneity exists on all spatial scales in the diffuse ISM
\citep[\eg,][]{Armstrong1995,Lazarian2000,Chepurnov2010},
and the column density of H{\sc i} gas likely varies with
the order of unity \citep[\eg,][]{Burkhart2015,Fukui2018}.
We also investigate the conditions with extremely low levels of fluctuation (as
low as $0.1$ \%) to understand the dependence of $\mydrhop$.
We choose the spatial resolution of $\Delta x=$
$7.8 \times 10^{-2}$, $3.9 \times 10^{-2}$, $2.0 \times 10^{-2}$
and $9.8 \times 10^{-3}$ pc by splitting the calculation domain with
from 256 $\times$ 256 $\times$ 128 cells at the lowest resolution
up to 2048 $\times$ 1024 $\times$ 1024 cells at the highest resolution
(hereafter noted as $\Delta x =0.08$ pc, $0.04$ pc, $0.02$ pc, and $0.01$ pc for simplicity
in the text and figures).
In addition, we repeat simulations with the same $\mydrhop$ and $\Delta x$
but with three different random phases $\myaa$, $\myab$, and $\myac$ to investigate
the possible range over which the averaged properties vary
due to such randomness even under the same statistical condition.
Note that the highest resolution is computationally expensive
and is limited to study in 3 representative cases only
($\mydrhop=31.6$, $10$, and $3.16$ \% with Phase $\myaa$).
\begin{figure*}
    {\large (a) $\mydrhop =$ 100\% } 
    \hspace{6.0cm} {\large (d) $\mydrhop =$ 3.16\%}  \\
    \begin{minipage}[t]{0.47\textwidth}
        \centering{\includegraphics[scale=0.5]{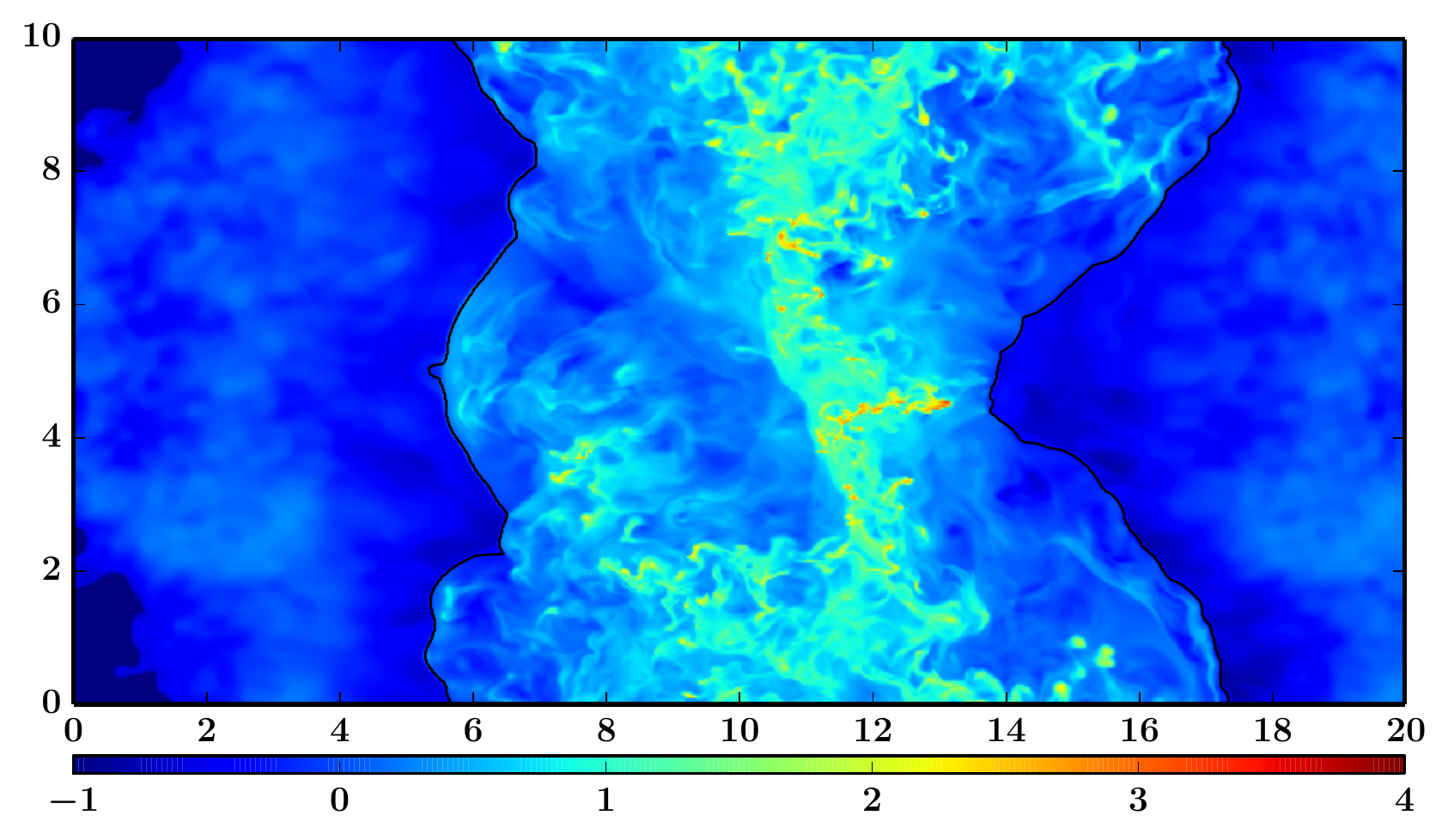}}
    \end{minipage}
    \hspace{0.5cm}
    \begin{minipage}[t]{0.47\textwidth}
        \centering{\includegraphics[scale=0.5]{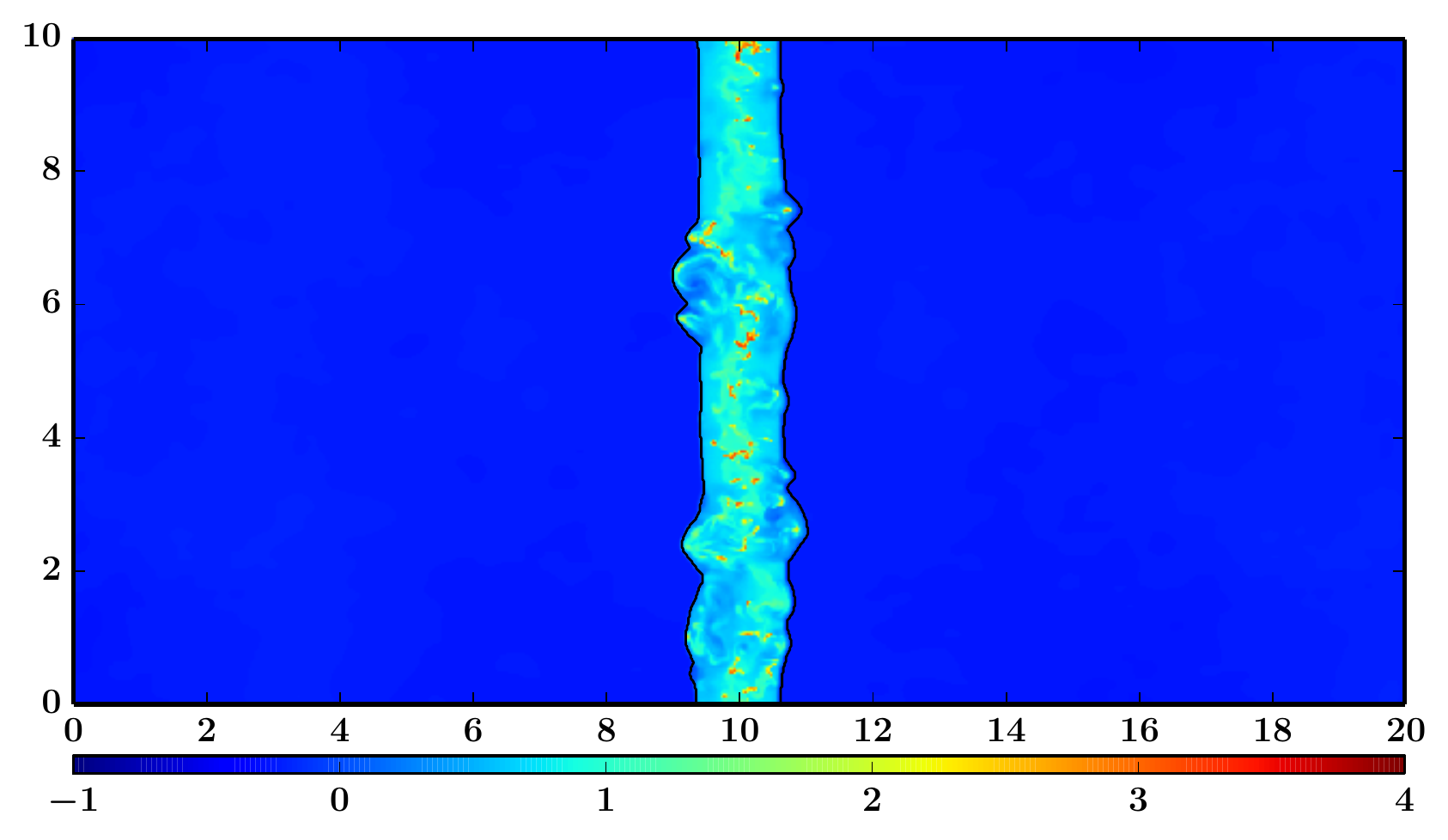}}
    \end{minipage}\\
    \hspace{1.0cm} {\large (b) $\mydrhop =$ 31.6\%}
    \hspace{6.0cm} {\large (e) $\mydrhop =$ 1\%}  \\
    \begin{minipage}[t]{0.47\textwidth}
        \centering{\includegraphics[scale=0.5]{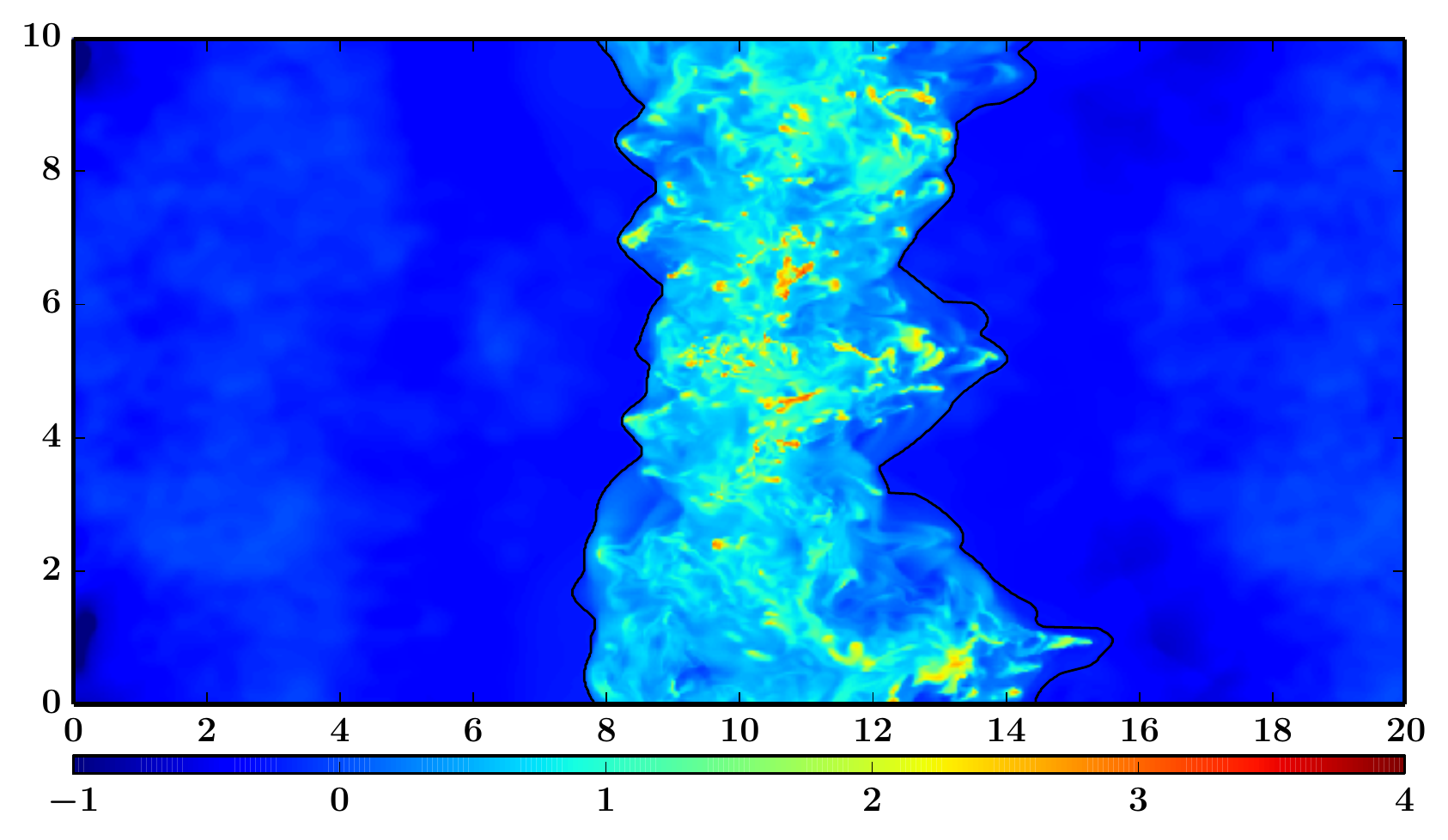}}
    \end{minipage}
    \hspace{0.5cm}
    \begin{minipage}[t]{0.47\textwidth}
        \centering{\includegraphics[scale=0.5]{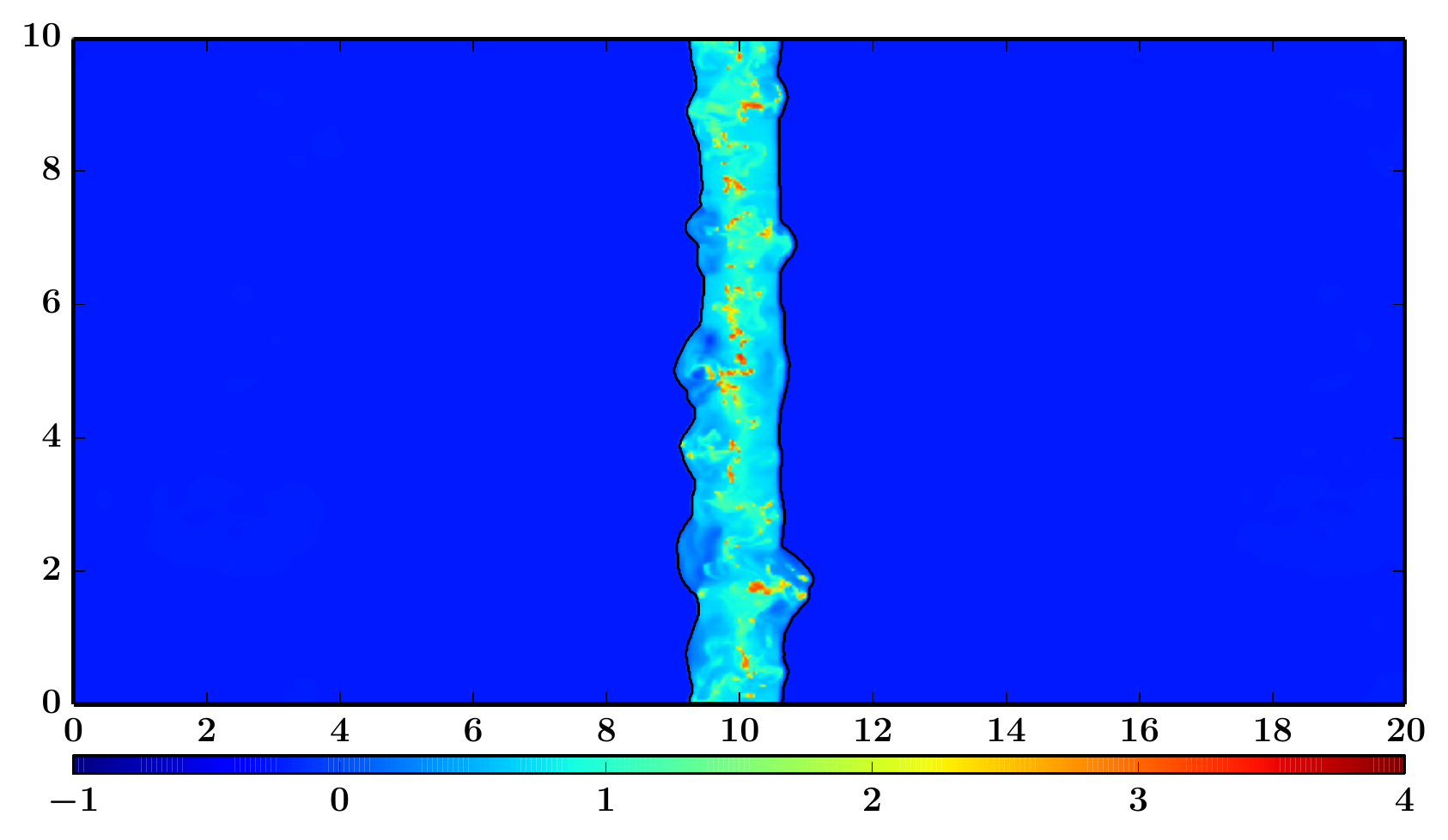}}
    \end{minipage}\\
    \hspace{1.0cm} {\large (c) $\mydrhop =$ 10\%}
    \hspace{6.4cm} {\large (f) $\mydrhop =$ 0\%}  \\
    \begin{minipage}[t]{0.47\textwidth}
        \centering{\includegraphics[scale=0.5]{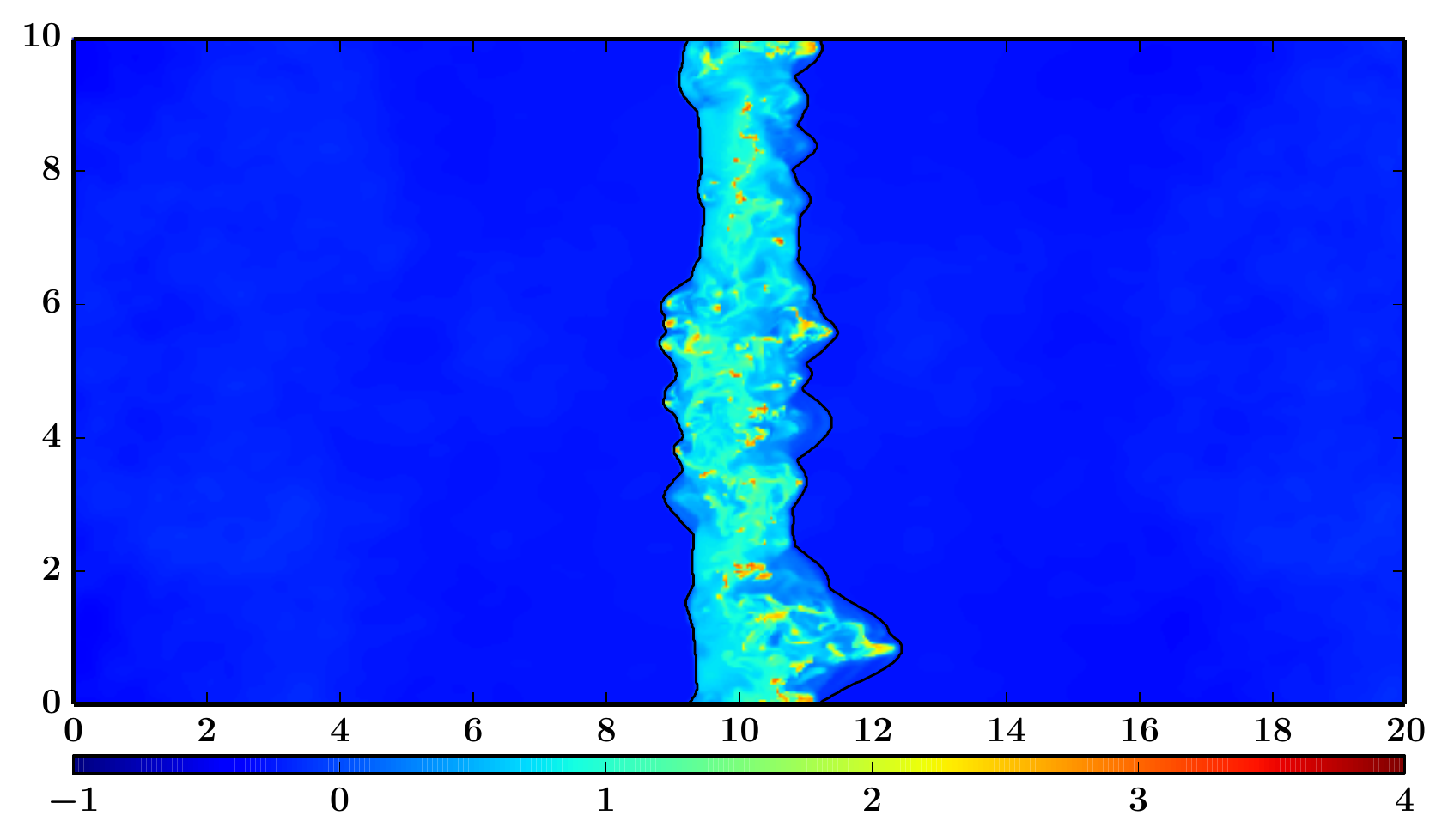}}
    \end{minipage}
    \hspace{0.5cm}
    \begin{minipage}[t]{0.47\textwidth}
        \centering{\includegraphics[scale=0.5]{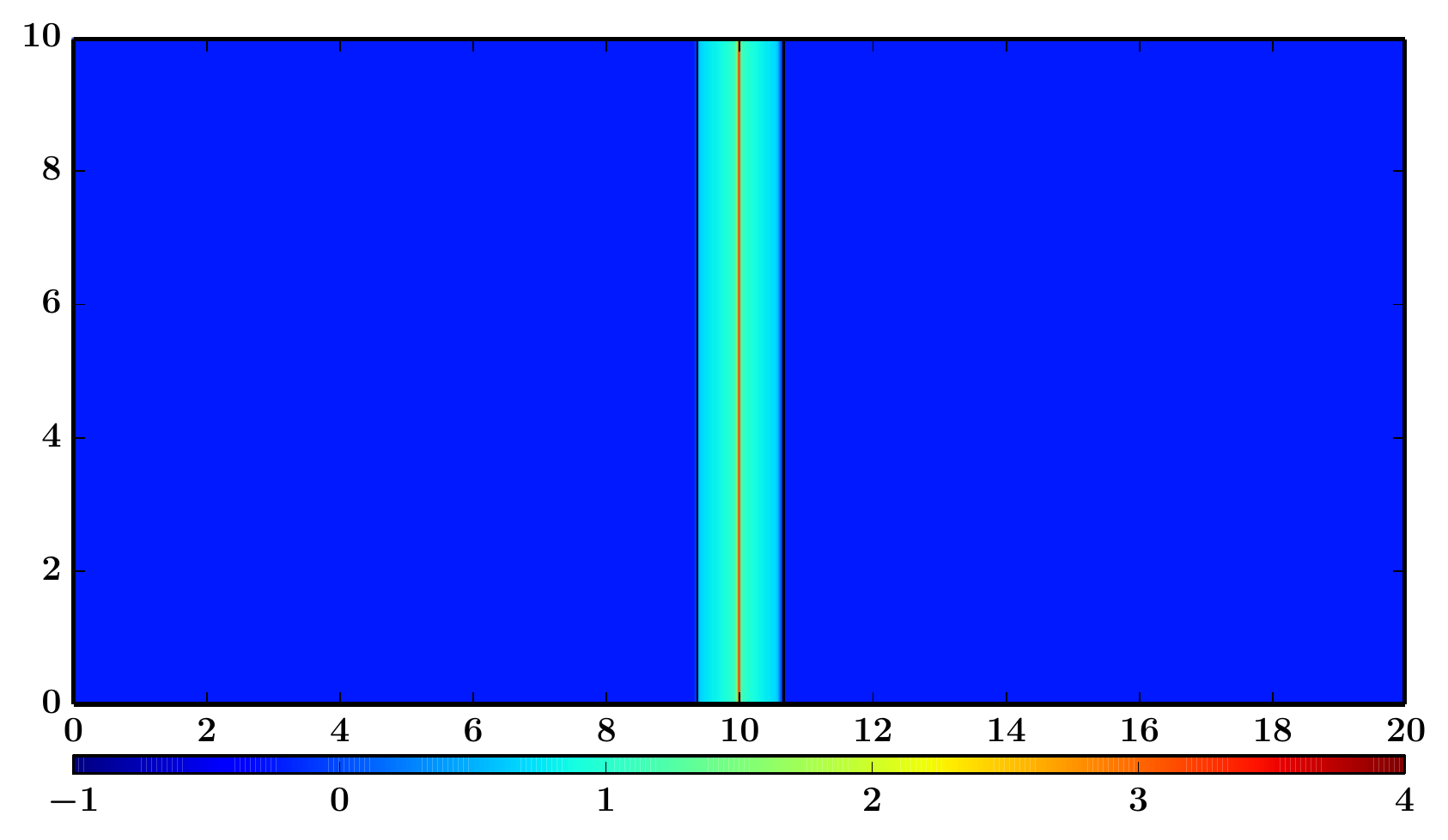}}
    \end{minipage}\\
    \vspace{-15pt}
    \caption{The density slice on $z=5$ pc plane at $1.8$ Myr,
             where the color represents $\log(n[{\rm cm^{-3}}])$.
             The horizontal and vertical axes are $x$ and $y$ directions
             in the unit of pc.
             The thin black lines show the shock front positions.
             The panels correspond to (a) $\mydrhop =$ 100\% (Run C2D1000), 
             (b) 31.6\% (Run C2D0316), (c) 10\% (Run C2D0100), (d) 3.16\% (Run C2D0031), 
             (e) 1\% (Run C2D0010), and (f) 0 \% (Run C2D0000) with Phase $\myaa$.
             Note that small-scale details are smeared out on these figures due to rasterization.
    }
    \label{fig:f3f4f5f6f7f8}
\end{figure*}

The combination of these 3(+1) resolutions, 
7 fluctuation amplitudes, and 3 phases
corresponds to 66 cases.
We additionally perform 3 controlled runs as a reference,
where we investigate head-on collisions
with the upstream WNM density completely uniform ($\mydrhop=0$ \%).
We thus investigate 69 cases in total.
Table~\ref{table:cases} summarizes the parameter sets, %that we systematically investigated,
where we list 45 cases that we will present in this article, out of the total 69 cases.
The calculation results are sampled every 0.1 Myr in each run.

\section{Results}
\label{sec:results}
\subsection{General Outcomes}
\label{subsec:genout}
In this section, we provide a brief overview of our simulation results,
showing how the geometrical structure of the shock-compressed layer varies
with $\mydrhop$ and cooling process.
Figure~\ref{fig:f3f4f5f6f7f8} shows the gallery of the density slice 
from runs with $\mydrhop =$ 100\% (Run C2D1000), 
31.6\% (Run C2D0316), 10\% (Run C2D0100), 3.16\% (Run C2D0031), 
1\% (Run C2D0010), and 0 \% (Run C2D0000) with Phase $\myaa$
and $\Delta x  = 0.02$ pc.
The shock-compressed layer is wider and wound more significantly 
with larger $\mydrhop$ due to the larger density fluctuation,
whereas the layer is narrower and tends to be less deformed with smaller $\mydrhop$.
Non-zero $\mydrhop$ provides the interaction between shocks and density inhomogeneity,
which drives turbulence
in the shock-compressed layer.
In contrast, $\mydrhop=0$ \% is an extreme condition where the flow becomes one-dimensional 
(\ie, the layer maintains a completely straight geometry), 
and all the mass of the WNM flow cools down and eventually accretes onto the thin CNM sheet formed at the center.

\begin{figure}\centering{
\includegraphics[width=1.0\columnwidth,keepaspectratio]{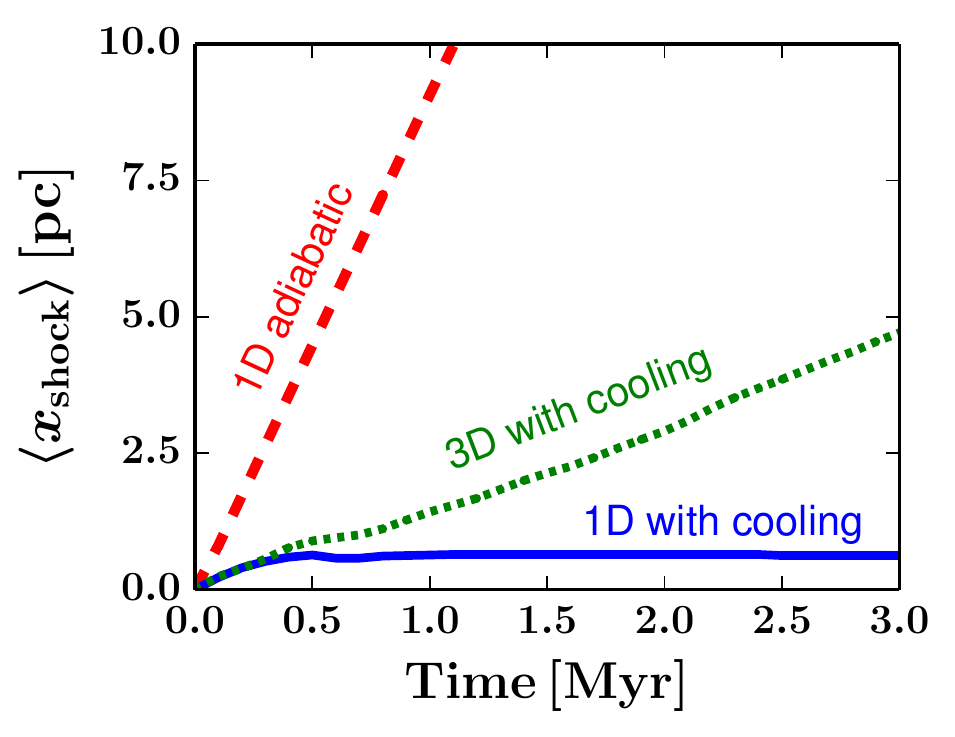}}
\caption{The time evolution of $\myxshock$ from three runs:
    an adiabatic flow with $\mydrhop=0$ \% (Run A2D0000, red dasehd line ``1D adiabatic''),
    a flow with cooling with $\mydrhop=0$ \% (Run C2D0000, blue solid line ``1D with cooling''),
    and a flow with cooling with $\mydrhop=31.6$ \% with Phase $\myaa$ 
    (Run C2D0316, green dotted line ``3D with cooling'').
    }
\label{fig:f9}
\end{figure}

Figure~\ref{fig:f9} shows the time evolution of $\myxshock$
from three runs: an adiabatic flow with $\mydrhop=0$ \% (Run A2D0000, ``1D adiabatic''),
a flow with cooling with $\mydrhop=0$ \% (Run C2D0000, ``1D with cooling''),
and a flow with cooling with $\mydrhop=31.6$ \% with Phase $\myaa$ (Run C2D0316, ``3D with cooling'').
As seen from the difference between Runs A2D0000 and C2D0000,
the cooling process transforms the WNM to the CNM
and the shock-compressed layer becomes denser and narrower.
As shown in Run C2D0316, 
a non-zero $\mydrhop$ converts a fraction of the WNM
to the CNM and drive some turbulence,
which makes the shock-compressed layer less dense and wider
compared with the uniform case (Run C2D0000),
but still significantly denser and narrower than 
the adiabatic case (Run A2D0000).
All the shock-compressed layer with non-zero $\mydrhop$
in our simulations
evolves somewhere between Runs A2D0000 and C2D0000,
accordingly.

\subsection{Convergence with Respect to the Spatial Resolution}
\label{subsec:conv}
\begin{figure*}
    \hspace{1.0cm}{\large (a) $\mydrhop =$ 31.6\% } 
    \hspace{6.0cm} {\large (b) $\mydrhop =$ 3.16\%}  \\
    \begin{minipage}[t]{0.47\textwidth}
        \centering{\includegraphics[scale=0.9]{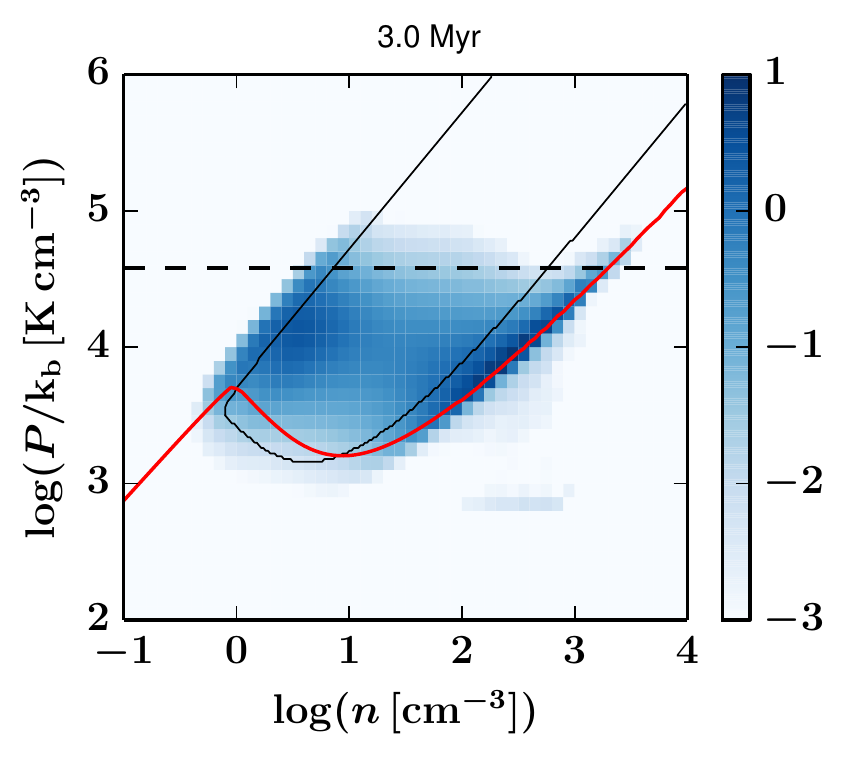}}
    \end{minipage}
    \hspace{0.5cm}
    \begin{minipage}[t]{0.47\textwidth}
        \centering{\includegraphics[scale=0.9]{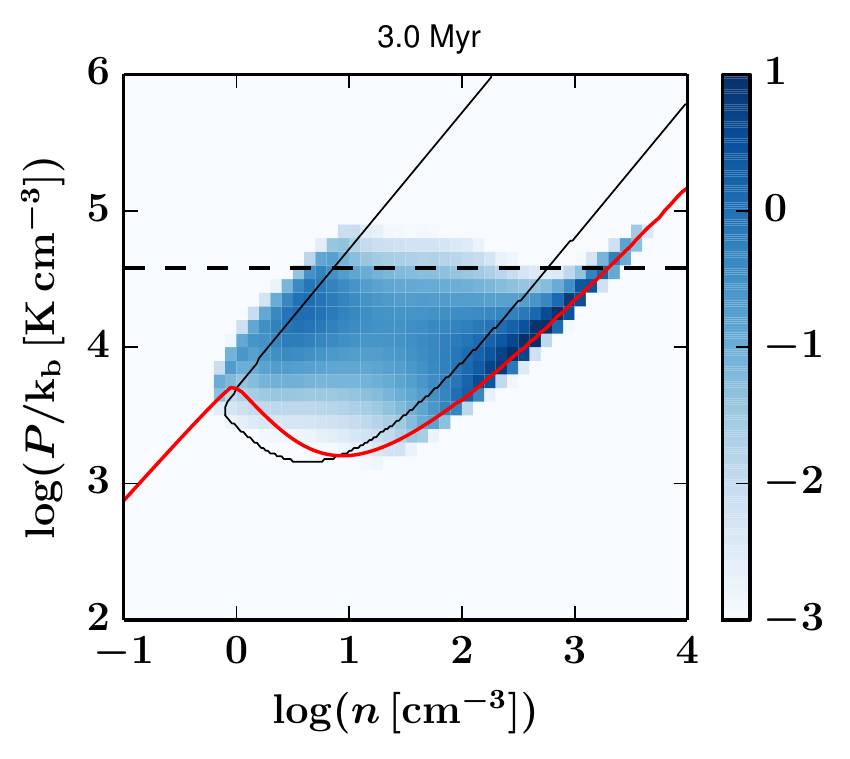}}
    \end{minipage}\\

    \hspace{1.0cm}{\large (c) $\mydrhop =$ 31.6\% } 
    \hspace{6.0cm} {\large (d) $\mydrhop =$ 3.16\%}  \\
    \begin{minipage}[t]{0.47\textwidth}
        \centering{\includegraphics[scale=0.75]{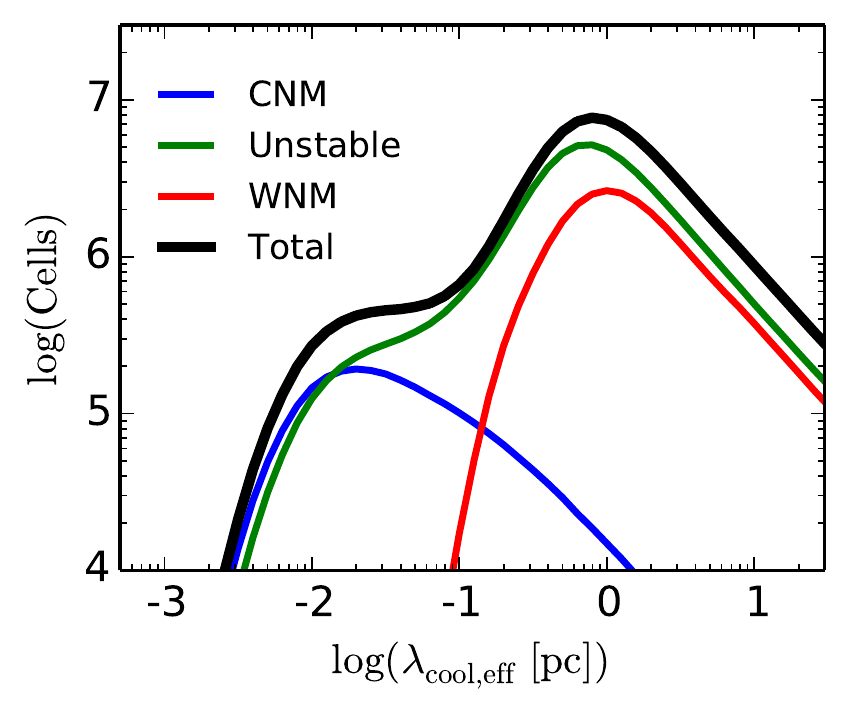}}
    \end{minipage}
    \hspace{0.5cm}
    \begin{minipage}[t]{0.47\textwidth}
        \centering{\includegraphics[scale=0.75]{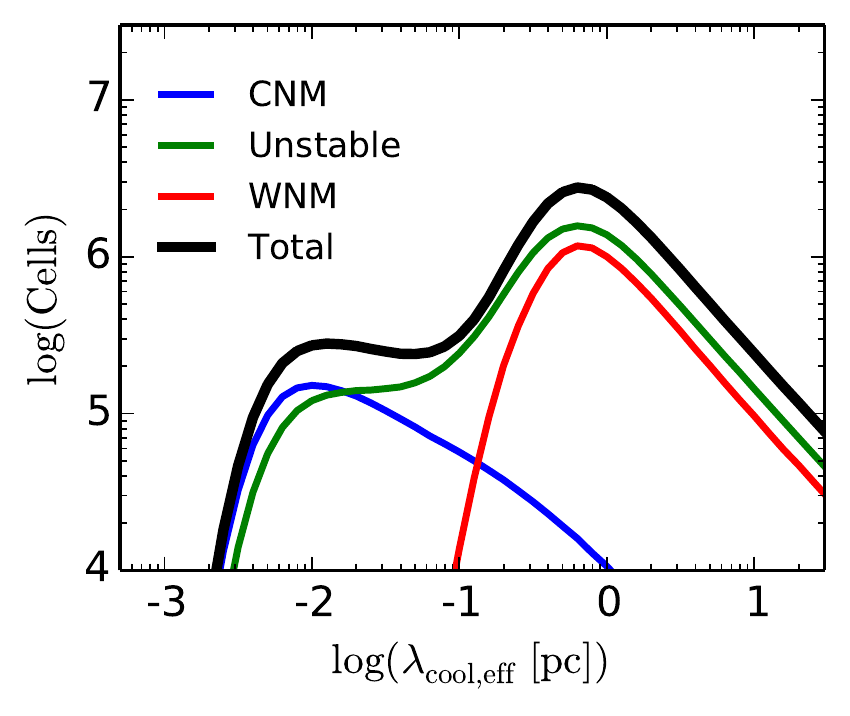}}
    \end{minipage}\\
    \vspace{-5pt}
    \caption{Panels (a) and (b): The mass histogram on the $P$-$n$ diagram at 3 Myr,
             where $\mydrhop=31.6$ \% (Run C2D0316, Panel (a)) and 
             $3.16$ \% (Run C2D0031, Panel (b)), both of which employ Phase $\myaa$ and $\Delta x =0.02$ pc.
             The blue color corresponds to $\log($mass$ [\msun])$.
             The red solid curve shows the thermal equilibrium region where the heating and cooling balances each other 
             defined as $\rho \myCooL = 0$.
             The horizontal black dashed line shows the ram pressure of the flow $\rho_0 \myvin^2$.
             The thin black curve encloses the region of the UNM defined as $(\partial (\mathcal{L}/T) / \partial T)_P <0$.
             The horizontal and vertical axes are logarithmically equally binned by a factor $1.26$.
             Panels (c) and (d): The histogram of the effective cooling length 
             $\lambda_{\rm cool,eff}$
             %$\mylcool$ 
             in the shock-compressed layer at 3 Myr,
             where again $\mydrhop=31.6$ \% (Run C2D0316, Panel (c)) and $3.16$ \% (Run C2D0031, Panel (d)).
             The blue curve shows the CNM, the green for the UNM, and red for the WNM,
             whereas the black curve shows the total of these components.
             %The vertical red line shows the mean cooling length $\langle \mylcool \rangle$: $2.3$ pc (Panel c) and $1.5$ pc (Panel d).
             %The bimodality in this histogram corresponds to the WNM component ($\mylcool \simeq 2$ pc)
             %and the CNM component ($\mylcool \simeq 0.002$ pc).
             The horizontal axis is logarithmically equally binned by a factor $1.26$.
             }
    \label{fig:f10f11f12f13}
\end{figure*}

\begin{figure*}
    {\large (a) $\mydrhop =$ 31.6 \%} 
    \hspace{6.0cm} {\large (b) $\mydrhop =$ 3.16 \%}  \\
    \begin{minipage}[t]{0.47\textwidth}
        \centering{\includegraphics[scale=0.8]{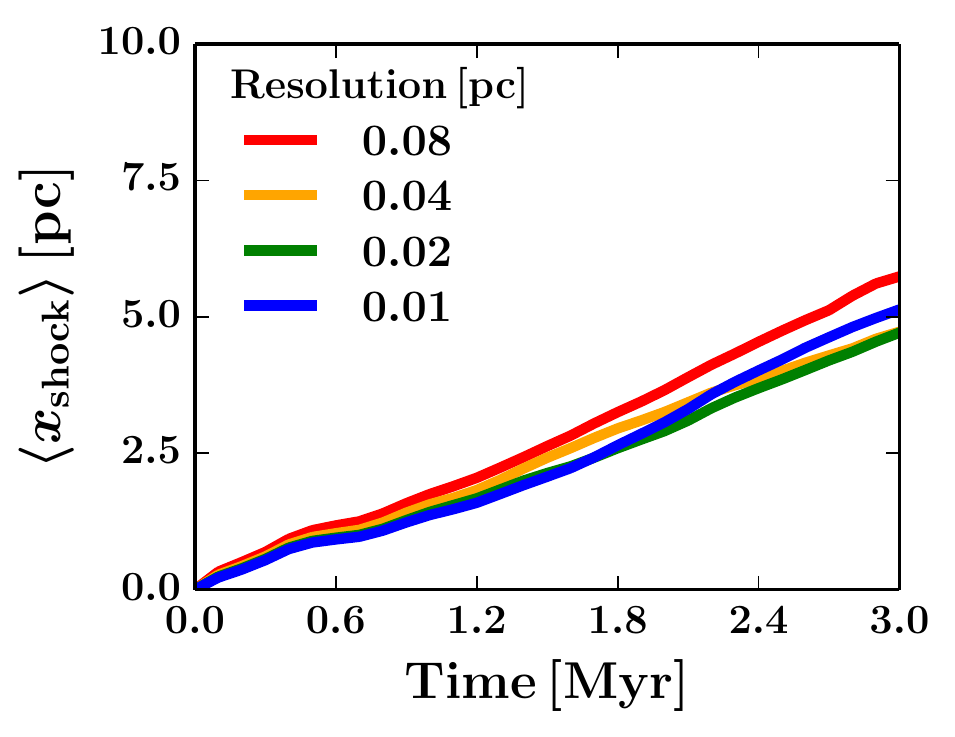}}
    \end{minipage}
    \hspace{0.5cm}
    \begin{minipage}[t]{0.47\textwidth}
        \centering{\includegraphics[scale=0.8]{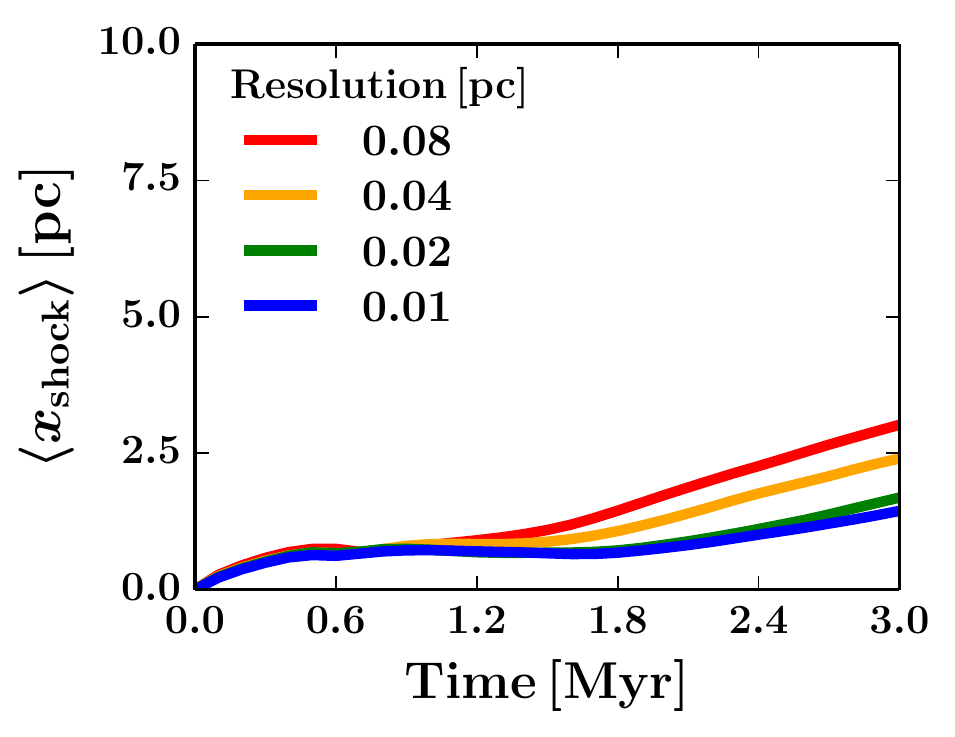}}
    \end{minipage}\\
    {\large (c) $\mydrhop =$ 31.6\% } 
    \hspace{6.0cm} {\large (d) $\mydrhop =$ 3.16\%}  \\
    \begin{minipage}[t]{0.47\textwidth}
        \centering{\includegraphics[scale=0.8]{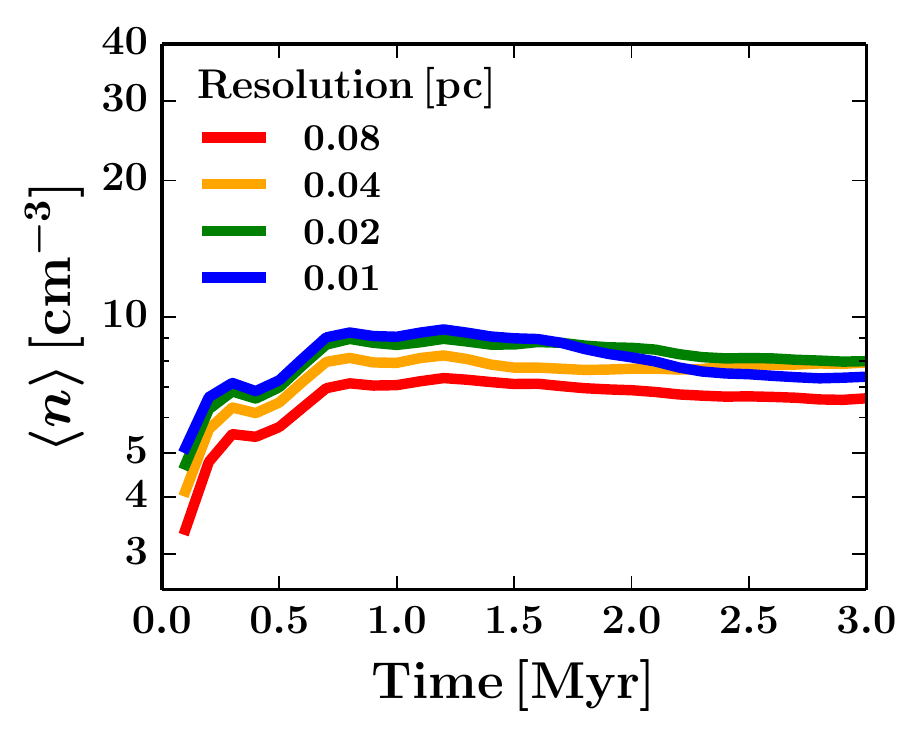}}
    \end{minipage}
    \hspace{0.5cm}
    \begin{minipage}[t]{0.47\textwidth}
        \centering{\includegraphics[scale=0.8]{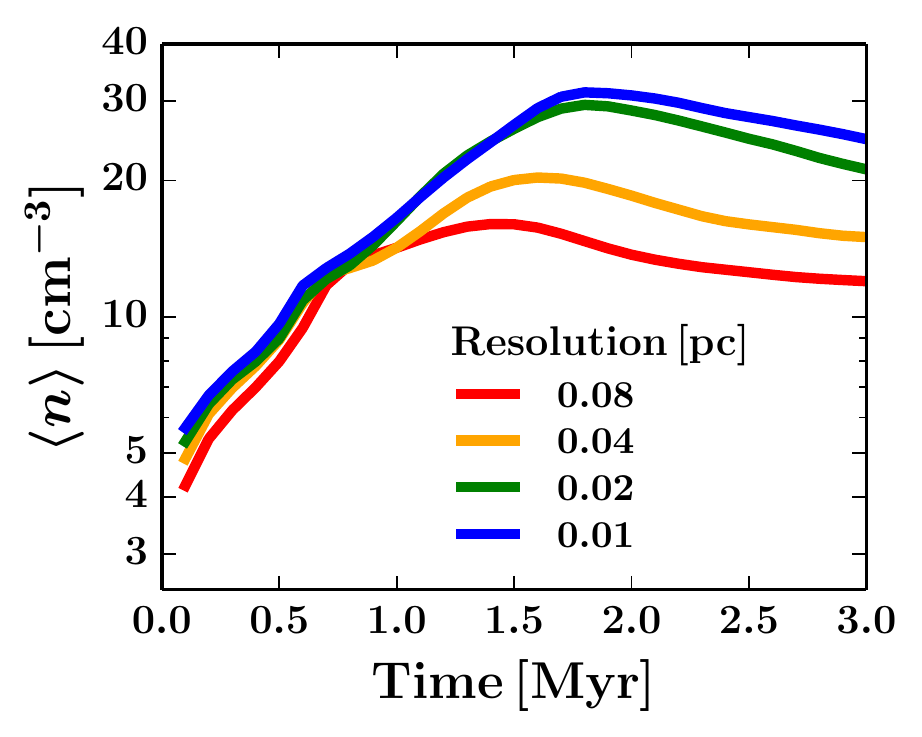}}
    \end{minipage}\\
    \vspace{-5pt}
    \caption{Panels (a) and (b): The time evolution of the average shock front position $\myxshock$ as a function of the spatial resolution $\Delta x$,
             where $\mydrhop=31.6$ \% (Runs C*D0316, Panel (a)) and $3.16$ \% (Runs C*D0031, Panel (b)), both of which employ Phase $\myaa$.
             The color corresponds to the spatial resolution.
             Panels (c) and (d): Same as Panels (a) and (b) but shows the time evolution of the mean density $\langle n \rangle$.
             }
    \label{fig:f14f15f16f17}
\end{figure*}
\begin{figure*}
    {\large (a) $\mydrhop =$ 31.6\% } 
    \hspace{6.0cm} {\large (b) $\mydrhop =$ 3.16\%}  \\
    \begin{minipage}[t]{0.47\textwidth}
        \centering{\includegraphics[scale=0.8]{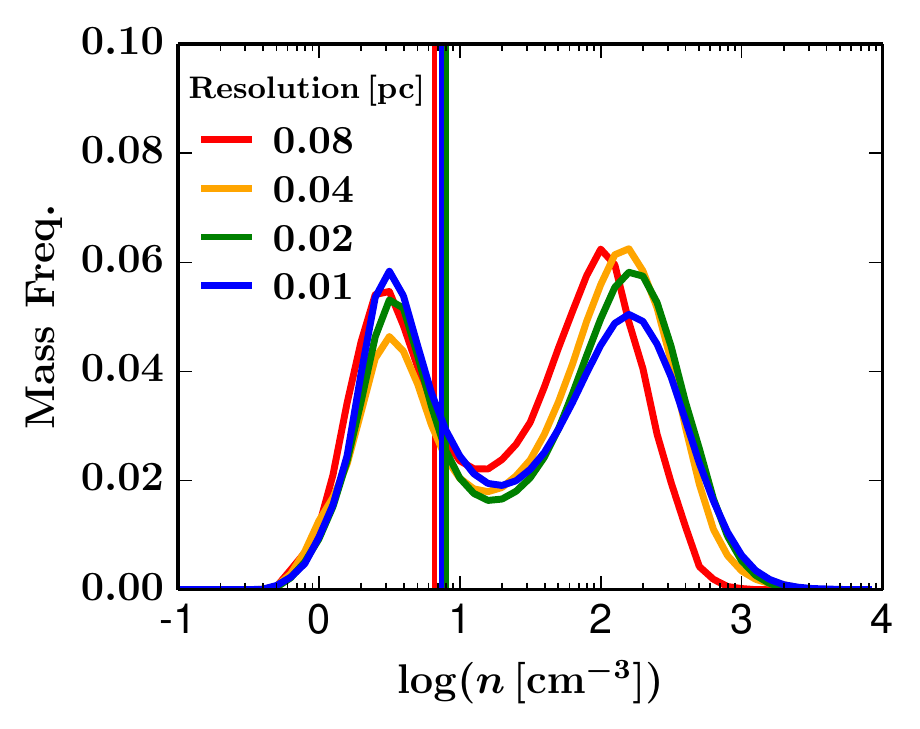}}
    \end{minipage}
    \hspace{0.5cm}
    \begin{minipage}[t]{0.47\textwidth}
        \centering{\includegraphics[scale=0.8]{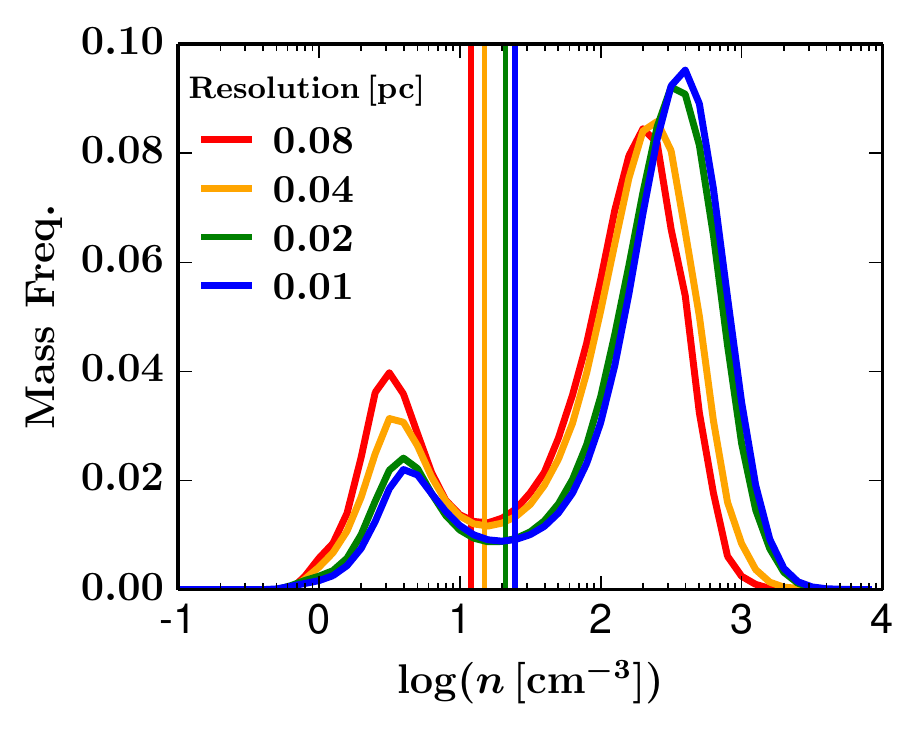}}
    \end{minipage}\\
    {\large (c) Various $\mydrhop$ with $\Delta x = 0.02$ pc} 
    \hspace{4.0cm} {\large (d) Variation with respect to $\Delta x = 0.01$ pc}  \\
    \begin{minipage}[t]{0.47\textwidth}
        \centering{\includegraphics[scale=0.8]{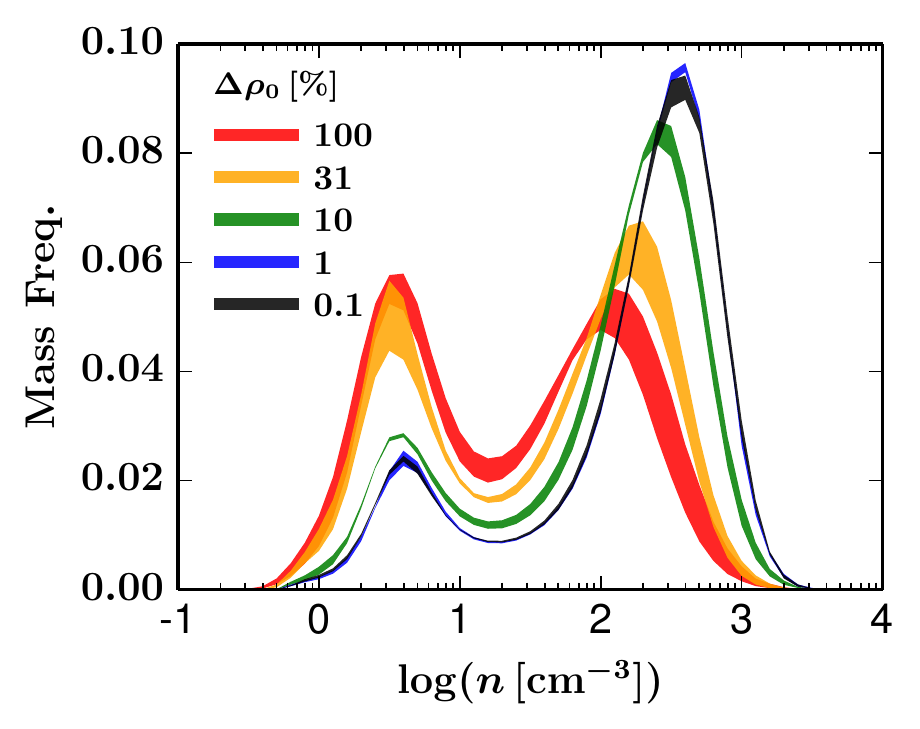}}
    \end{minipage}
    \hspace{0.5cm}
    \begin{minipage}[t]{0.47\textwidth}
        \centering{\includegraphics[scale=0.8]{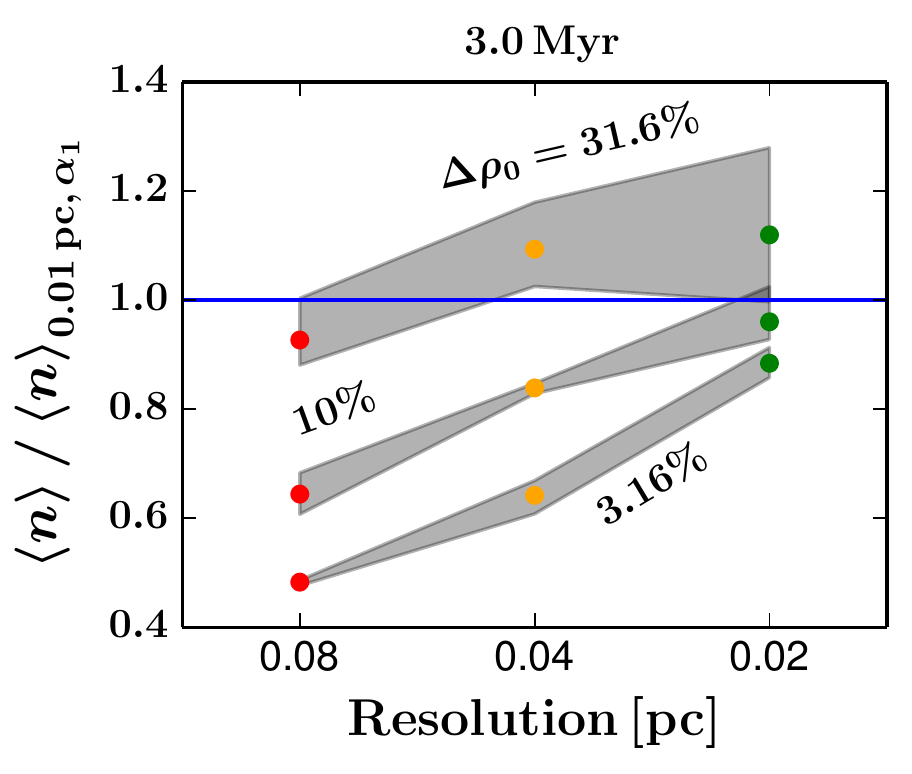}}
    \end{minipage}\\
    \vspace{-5pt}
    \caption{Panels (a) and (b): The mass frequency ${\rm d}M/{\rm d}\log(n)$ as a function of $n$ at 3 Myr
             with $\mydrhop=31.6$ \% (Runs C*D0316, Panel (a)) and $3.16$ \% (Runs C*D0031, Panel (b)), 
             both of which employ Phase $\myaa$.
             The color corresponds to the spatial resolution $\Delta x$.
             The vertical lines show the mean density $\langle n \rangle$, whose color coding is matched with each spatial resolution.
             Panel (c): Compilation of the mass frequency from $\mydrhop =$ 100\% (Run C2D1000), 
             31.6\% (Run C2D0316), 10\% (Run C2D0100), 1\% (Run C2D0010), and 0.1 \% (Run C2D0001).
             The shades in each $\mydrhop$ show the range of the maximum and minimum due to Phases $\myaa$, $\myab$, and $\myac$.
             Note that the shades of $\mydrhop=1$ \% and $0.1$ \% are difficult to read because they are almost overlapped.
             Panel (d): the ratio of $\langle n \rangle (\Delta x, \alpha)$ to the $\langle n \rangle (0.01\,\mathrm{pc}, \myaa)$
             in each $\mydrhop$.
             %$\langle n \rangle$ as a function of $\Delta x$ with respect to the result from $\Delta x = 0.01$ pc in each $\mydrhop$.
             The grey shades correspond to the range of the maximum and minimum due to Phases $\myaa$, $\myab$, and $\myac$.
             The colored points correspond to the average within each combination of ($\mydrhop$, $\Delta x$),
             where the colors correspond to $\Delta x=0.08$ pc (red), $0.04$ pc (orange), and $0.02$ pc (green).
             Note that runs with $\Delta x = 0.01$ pc are limited to Phase $\myaa$, and therfore the denominator in the vertical axis
             is a single value $\langle n \rangle$ from the Phase $\myaa$ but not the average of the three phases.
             }
    \label{fig:f18f19f20f21}
\end{figure*}

We first investigate the convergence with respect to the spatial resolution $\Delta x$
by varying it from $0.08$ pc to $0.01$ pc in all $\mydrhop$ cases
as shown in Table~\ref{table:cases}.
We find that the mean properties of the shock-compressed layer is converged 
with $0.02$ pc spatial resolution in any $\mydrhop$ case, 
but the trend of convergence differs between the cases with larger $\mydrhop > 10$ \%
and with smaller $\mydrhop \leq 10$ \%.
Therefore in this section, we show the results mainly from $\mydrhop=31.6$ \%  and $3.16$ \% as 
examples of large and small $\mydrhop$ and investigate how the physical properties of the shock-compressed layer converge.

Panels (a) and (b) in Figure~\ref{fig:f10f11f12f13} show the mass histogram 
on the $P$-$n$ diagram at 3 Myr from $\mydrhop=31.6$ \% (Run C2D0316, Panel (a)) and $3.16$ \% (Run C2D0031, Panel (b)).
Most mass resides in two regions: the shock-heated component (WNM and UNM at $n \simeq 2 \cmkk$) and 
the cooled component (CNM at $n > 100 \cmkk$).
As shown here, the WNM density is typically 100 times diffuse than that of CNM,
and the volume of the shock-compressed layer is dominated by the WNM\@.
Therefore, resolving the typical transition scale from the WNM into CNM 
is crucial for the convergence in the mean properties 
of the shock-compressed layer,
while resolving the CNM cooling length is important when we investigate the detailed structures of CNM clumps.
We thus investigate the effective cooling length $\lambda_{\rm cool,eff}$ in the shock-compressed layer at 3 Myr.
Panels (c) and (d) in Figure~\ref{fig:f10f11f12f13}
show its histogram.
Here we calculate $\lambda_{\rm cool,eff}$ as $\mycs P/(\gamma-1)/(n^2 \Lambda - n \Gamma + P \nabla_{\mu} v_{\mu})$ 
in each cell where $\mycs$ and $P$ are the 
sound speed and thermal pressure in each cell.
The panels show that $0.02$ pc corresponds to the typical spatial scale
on which the dominant components in the cooling length transits from the UNM to CNM.
We also find that the spatial resolution of $0.02$ pc fully resolves the WNM cooling length,
and resolves the cooling length of more than 99 \% (92 \%) of the UNM in volume (mass)
when $\mydrhop=31.6$ \% and more than 98 \% (80 \%) when $\mydrhop=3.16$ \%.
We thus expect that the mean properties 
converge when $\Delta x = 0.02{\rm~pc}$ or higher resolution,
with which we can follow the dynamical condensation from WNM and UNM to CNM due to cooling.

This characteristic scale of $0.02$ pc is also consistent with the resolution requirement empirically suggested by 
previous numerical simulations. For example, 
\cite{Inoue2015} performed converging flow calculations similar to our studies\footnote[10]{\cite{Inoue2015}
start with the UNM
%thermally unstable medium 
with $\langle n \rangle=2.5 \, \cmkk$.},
and suggested that more than 60 cells on 
the most frequent cooling length
(defined as $\mylcool = \mycs P/(\gamma-1)/(n^2 \Lambda)$)
%a typical cooling scale
is required to have convergence in the mass probability distribution function
and CNM clump mass function.
In our simulations, 
the mean cooling scale $\langle \mylcool \rangle$ is peaked at the WNM and UNM component with 
$3.4$ pc in $\mydrhop=31.6$ \% and $2.2$ pc in $3.16$ \%.
The requirement of more than 60 cells over one cooling scale corresponds to
$\Delta x < 0.037$ pc, accordingly.
%\cite{Inoue2015} perform converging flow calculations similar to our studies\footnote[10]{\cite{Inoue2015}
%start with thermally unstable medium with $\langle n \rangle=2.5 \, \cmkk$.},
%and suggest that more than 60 cells on a typical cooling scale
%is required to have convergence in the mass probability distribution function
%and CNM clump mass function.
%We thus investigate the cooling length $\mylcool$ in the shock-compressed layer at 3 Myr,
%and Panels (c) and (d) in Figure~\ref{fig:f10f11f12f13}
%show its histogram.
%Here we calculate $\lambda_{\rm cool}$ as $\mycs P/(n^2 \Lambda)$ in each cell where $\mycs$ and $P$ are the 
%sound speed and thermal pressure in each cell.
%The bimodality in this histogram corresponds to the WNM component ($\mylcool \simeq 2$ pc)
%and the CNM component ($\mylcool \simeq 0.002$ pc).
%The mean cooling scale is peaked at the WNM component with 
%$\langle \mylcool \rangle = 2.3$ pc in $\mydrhop=31.6$ \% and $1.5$ pc in $3.16$ \%.
%The requirement of more than 60 cells over one cooling scale corresponds to
%$\Delta x < 0.025$ pc. 
%We thus expect that the mean properties 
%converge with $\Delta x = 0.02{\rm~pc}$ or higher in our simulations.

Panels (a) -- (d) in Figure~\ref{fig:f14f15f16f17} show 
such convergence in the mean properties;
the time evolution of $\myxshock$ and
$\langle n \rangle$ as a function of $\Delta x$ with Phase $\myaa$,
where Panels (a) and (c) for $\mydrhop=31.6$ \% and (b) and (d) for $3.16$ \%.
In both $\mydrhop$ cases, 
the linear growth of $\myxshock$ (\ie, the constant speed of the shock propagation)
corresponds to the quasi-steady $\langle n \rangle$.
As already seen in Figure~\ref{fig:f3f4f5f6f7f8},
the larger (smaller) $\mydrhop$ results in a wider (narrower) layer
and smaller (denser) $\langle n \rangle$.
The measured $\myxshock$ and $\langle n \rangle$ at 3 Myr are listed on
Table~\ref{table:resol}.
These results suggest that
$\myxshock$ and $\langle n \rangle$
indeed show a convergence with $\Delta x =0.02$ pc,
with less than 17 \%  difference between the results of $\Delta x =0.02$ pc
and $0.01$ pc.

The trend of convergence, however, differs between large and small $\mydrhop$.
For example, $\myxshock$ and $\langle n \rangle$ of Runs C*D0316 
do not vary monotonically along with $\Delta x$
whereas those of Runs C*D0031 do.
In addition, the difference in $\myxshock$ and $\langle n \rangle$ at 3 Myr 
between $\Delta x=0.08$ pc and $0.01$ pc 
is limited to a factor of $0.11$ in Runs C*D0316 
but is by a factor of $2.09$ in Runs C*D0031.
Thus overall, convergence with large $\mydrhop$ is non-monotonic 
and calculations with coarse resolutions of $\geq 0.02$ pc show similar values in 
$\myxshock$ and $\langle n \rangle$,
whereas the convergence with small $\mydrhop$ is monotonic and stringently requires
$0.02$ pc as expected from the cooling length analysis above.

To understand these trends, we investigate the mass frequency 
as a function of density within the shock-compressed layer,
which is shown in Panels (a) and (b) of Figure~\ref{fig:f18f19f20f21}
with $\langle n \rangle$ as vertical lines.
We also measure 
the mass fraction of WNM ($f_{\rm WNM}$), 
UNM ($f_{\rm UNM}$), and CNM ($f_{\rm CNM}$),
%the CNM mass fraction, $f_{\rm CNM}$,
%defined as \mk{$T<100\,\mathrm{K}$},
%$n\geq 10\, \cmkk$,
and the peak density of the CNM $n_{\rm CNM\, peak}$, which are summarized on Table~\ref{table:resol}.
The bimodality commonly appears in both $\mydrhop$ cases,
which shows the shock-heated component peaked at $n\sim2$ -- $4\,\cmkk$ mostly corresponding to the WNM and UNM,
and the cooled component peaked at $n\sim100$ -- $400\,\cmkk$ mostly corresponding to the CNM.
%The mass mainly resides in 
%WNM and CNM components in both $\mydrhop$ cases, 
%but their fraction differs. 
The relative mass fraction, however, depends on $\mydrhop$.
$f_{\rm CNM}$ remains $\sim 45$ \% when $\mydrhop=31.6$ \%
and both the WNM+UNM and CNM contribute to $\langle n \rangle$,
whereas $f_{\rm CNM}$ is $\sim 70$ \%  when $\mydrhop=3.16$ \%
and $\langle n \rangle$ depends more on the CNM component\footnote[11]{Given
the typical density difference between the WNM and CNM is $\sim 100$,
the volume ratio of 
WNM+UNM:CNM is $\sim 100:1$ when $\mydrhop=31.6$ \%
and $\sim 40:1$ when $\mydrhop=3.16$ \%.
%WNM:CNM is $\sim 67:1$ when $\mydrhop=31.6$ \%
%and $\sim 11:1$ when $\mydrhop=3.16$ \%.
Therefore, the shock-heated component (WNM+UNM)
%WNM component 
always dominates the volume
and the bimomdality already seen in Figure~\ref{fig:f10f11f12f13}
is a proxy of such volume ratio.
Also note that the choice of the WNM/CNM boundary has an arbitrariness
at $P/k_{\rm B} \sim 10^3$ K cm$^{-3}$ and $n\sim1\, \cmkk$.
However, 
the mass in such state (with $100\leq T <5000$ K and smaller pressure than that of the UNM) is limited to $<0.17$ \% ($<0.01$ \%) 
of the total mass when $\mydrhop=31.6$ \% (3.16 \%).
Therefore on Table~\ref{table:resol}, 
we count the mass with $T\geq1000$ K as WNM and $T<1000$ K as CNM
in that low-density low-pressure regime for simplicity.}.
%The WNM:CNM mass ratio is $\sim 4:6$ when $\mydrhop=31.6$ \%
%and both the WNM and CNM contribute to
%$\langle n \rangle$,
%whereas the ratio is $\sim 1:9$ when $\mydrhop=3.16$ \%
%and $\langle n \rangle$ depends more on the CNM component\footnote[11]{Given
%the typical density difference between the WNM and CNM is $\sim 100$,
%the volume ratio of WNM:CNM is $\sim 67:1$ when $\mydrhop=31.6$ \%
%and $\sim 11:1$ when $\mydrhop=3.16$ \%.
%Therefore, the WNM component always dominates the volume
%and the bimomdality already seen in Figure~\ref{fig:f10f11f12f13}
%is a proxy of such volume ratio.}.
For example, 
$\langle n \rangle$ with $\mydrhop=31.6$ \% shifts non-monotonically
in accordance with the non-monotonic change in the WNM fraction with $\Delta x$,
resulting in similar $\langle n \rangle$ between different $\Delta x$.
In contrast, $\langle n \rangle$ with $\mydrhop=3.16$ \% shifts monotonically by the factor
as $n_{\rm CNM, peak}$ monotonically shifts with $\Delta x$
(\eg, from $\Delta x=0.08$ pc to $0.01$ pc, $\langle n \rangle$ shifts
from $12.0$ to $24.7\, \cmkk$ with $n_{\rm CNM, peak}$ from $200$ to $400\, \cmkk$).
\begin{table*}
    \caption{Physical properties at 3 Myr
            from $\mydrhop=31.6$ \% (Runs C*D0316) and $3.16$ \% (Runs C*D0031) 
            with Phase $\myaa$, 
            with respect to the spatial resolution $\Delta x$.}
    \centering{
        \begin{tabular}{c||c|c}
            \hline
            \hline
            \multicolumn{1}{c||}{} & \multicolumn{2}{c}{$\myxshock$ [pc], 
            $\langle n \rangle$ [$\cmkk$],  
            $f_{\rm WNM}$ [\%], 
            $f_{\rm UNM}$ [\%], 
            $f_{\rm CNM}$ [\%], 
            $n_{\rm CNM\, peak}$ [$\cmkk$]} \\
            \hline 
            $\Delta x$ (pc) & $\mydrhop=31.6$ \%                                       &   $\mydrhop=3.16$ \%   \\ \hline 
 \hline
0.08   & \hspace{2pt} 5.74,\hspace{2pt} 6.61,\hspace{2pt} 9.7,\hspace{2pt} 52.0,\hspace{2pt} 38.3,\hspace{2pt} 100\hspace{4pt}  & 3.02,\hspace{2pt} 12.0,\hspace{2pt} 8.4,\hspace{2pt} 33.5,\hspace{2pt} 58.1,\hspace{2pt} 200\\
0.04   & \hspace{2pt} 4.73,\hspace{2pt} 7.93,\hspace{2pt} 8.6,\hspace{2pt} 44.2,\hspace{2pt} 47.2,\hspace{2pt} 158\hspace{4pt}  & 2.40,\hspace{2pt} 15.0,\hspace{2pt} 5.9,\hspace{2pt} 31.3,\hspace{2pt} 62.8,\hspace{2pt} 251\\
0.02   & \hspace{2pt} 4.71,\hspace{2pt} 7.99,\hspace{2pt} 7.9,\hspace{2pt} 45.3,\hspace{2pt} 46.8,\hspace{2pt} 158\hspace{4pt}  & 1.69,\hspace{2pt} 21.2,\hspace{2pt} 4.2,\hspace{2pt} 24.6,\hspace{2pt} 71.2,\hspace{2pt} 316\\
0.01   & \hspace{2pt} 5.14,\hspace{2pt} 7.38,\hspace{2pt} 7.6,\hspace{2pt} 49.9,\hspace{2pt} 42.5,\hspace{2pt} 158\hspace{4pt}  & 1.45,\hspace{2pt} 24.7,\hspace{2pt} 3.3,\hspace{2pt} 24.4,\hspace{2pt} 72.3,\hspace{2pt} 400\\

            \hline
            \hline
        \end{tabular}
    }\par
    \label{table:resol}
\end{table*}

Two distinct behaviors depending on $\mydrhop$ 
appear also with different random phases $\myaa$, $\myab$, and $\myac$
in the upstream density fluctuation.
Panel (c) of Figure~\ref{fig:f18f19f20f21} shows the 
mass frequency variation due to those three random phases
in each $\mydrhop$
under a fixed power spectrum of $P_{\rho}(k)\propto k^{-11/3}$
and a fixed spatial resolution of $\Delta x=0.02$ pc.
This shows that the mass frequency has a larger 
variation
by the random phases $\alpha$
in larger $\mydrhop$ cases.
To compare this variation due to different $\alpha$
with the variation due to $\Delta x$,
Panel (d) of Figure~\ref{fig:f18f19f20f21} shows the 
ratio of $\langle n \rangle (\Delta x, \alpha)$ 
to the $\langle n \rangle (0.01\,\mathrm{pc}, \myaa)$
in each $\mydrhop$.
In the case of $\mydrhop=31.6$ \%,
we found that 
the variation due to different $\alpha$
and that due to different $\Delta x$ are comparable
(\ie, the width of grey shades is comparable to the scatter
of points around the unity).
This suggests that resolving the density fluctuation with different $\Delta x$
also intrinsically produces the variation in $\langle n \rangle$
comparable to the variation produced by different $\alpha$.
Therfore, $\langle n \rangle$ 
converges non-monotonically 
and calculations with coarse resolutions of $\Delta x > 0.02$ pc 
practically provide simliar values in $\myxshock$ and $\langle n \rangle$
even though the physical convergence still requires $\Delta x=0.02$ pc.
The physical origin of such variation
due to different $\Delta x$ is explained by the 
non-linear nature of the system,
which we will discuss in Section~\ref{subsec:sto}.
In contrast in the case of $\mydrhop\leq10$ \%, we found that the variation due to different $\alpha$ is limited,
and the effect of changing $\Delta x$ directly appears as a monotonic change of $\langle n \rangle$,
showing the convergence 
as predicted by the cooling length analysis.

Therefore in the following sections, we will focus on the results based on $\Delta x = 0.02$ pc
and highlight how the physical properties of the shock-compressed layer varies 
with $\mydrhop$ and the random phases.

\begin{figure*}
    \begin{minipage}[t]{0.47\textwidth}
        \centering{\includegraphics[scale=0.9]{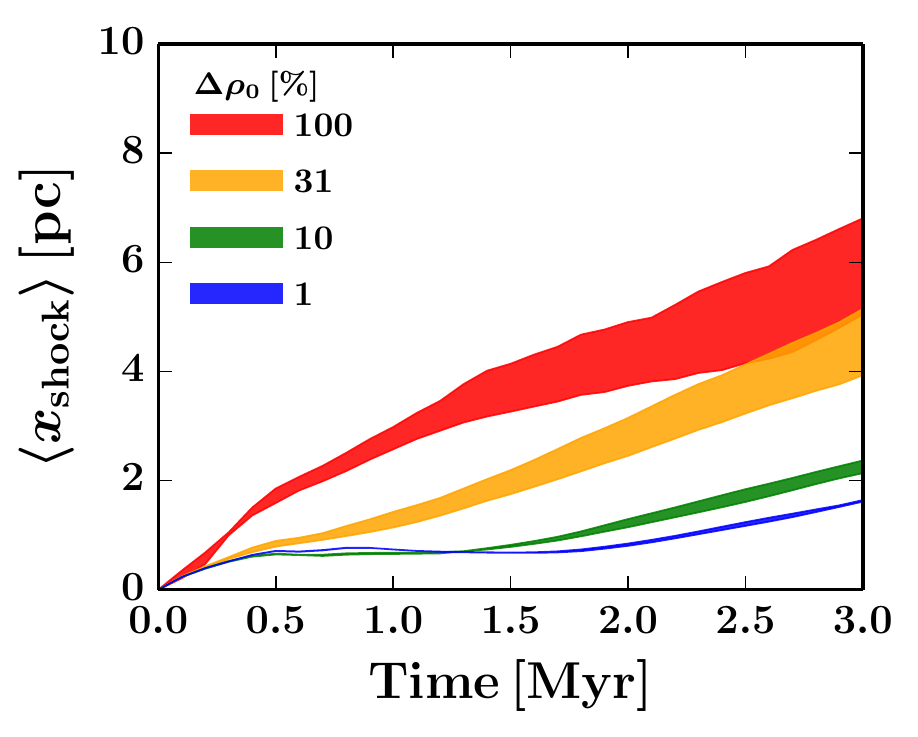}}
    \end{minipage}
    \hspace{0.5cm}
    \begin{minipage}[t]{0.47\textwidth}
        \centering{\includegraphics[scale=0.9]{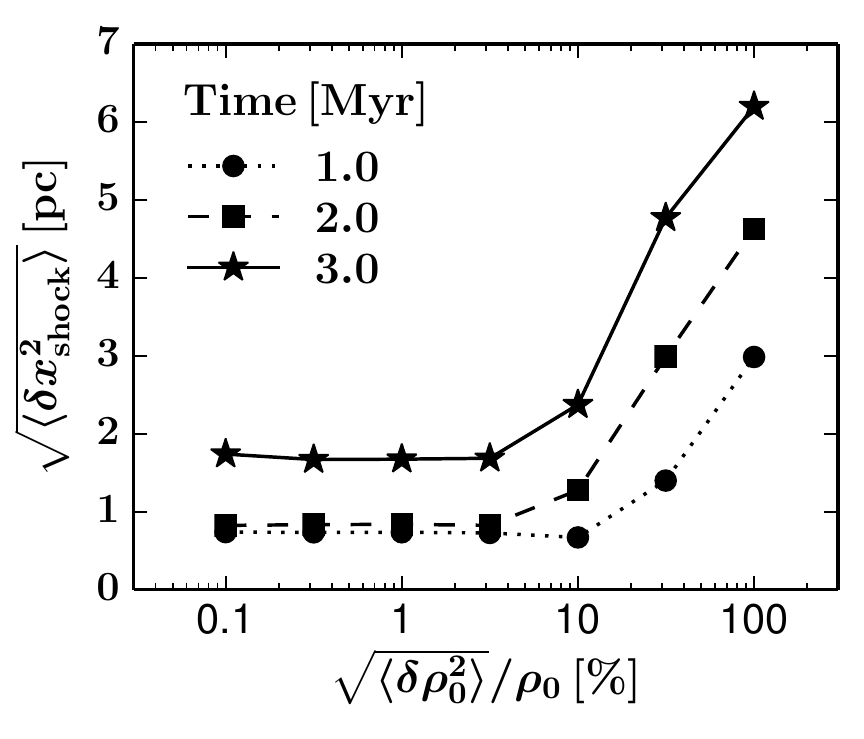}}
    \end{minipage}\\
    \caption{Left: The time evolution of the average shock front position $\myxshock$ as a function of $\mydrhop$
             with $\Delta x= 0.02$ pc.
             The color corresponds to $\mydrhop$. The shades in each $\mydrhop$ show the range 
             of the maximum and minimum due to Phase $\myaa$, $\myab$, and $\myac$.
             Right: the dispersion of the shock front position $\sqrt{\langle \delta x_{\rm shock}^2 \rangle}$
             as a function of $\mydrhop$ at 1, 2, and 3 Myr. We here average the results of three random phases at each time step.
    }
    \label{fig:f22f23}
\end{figure*}
\begin{figure*}
    {\large (a) $\mydrhop =$ 100\% } 
    \hspace{6.0cm} {\large (b) $\mydrhop =$ 1\%}  \\
    \begin{minipage}[t]{0.47\textwidth}
        \centering{\includegraphics[scale=0.45]{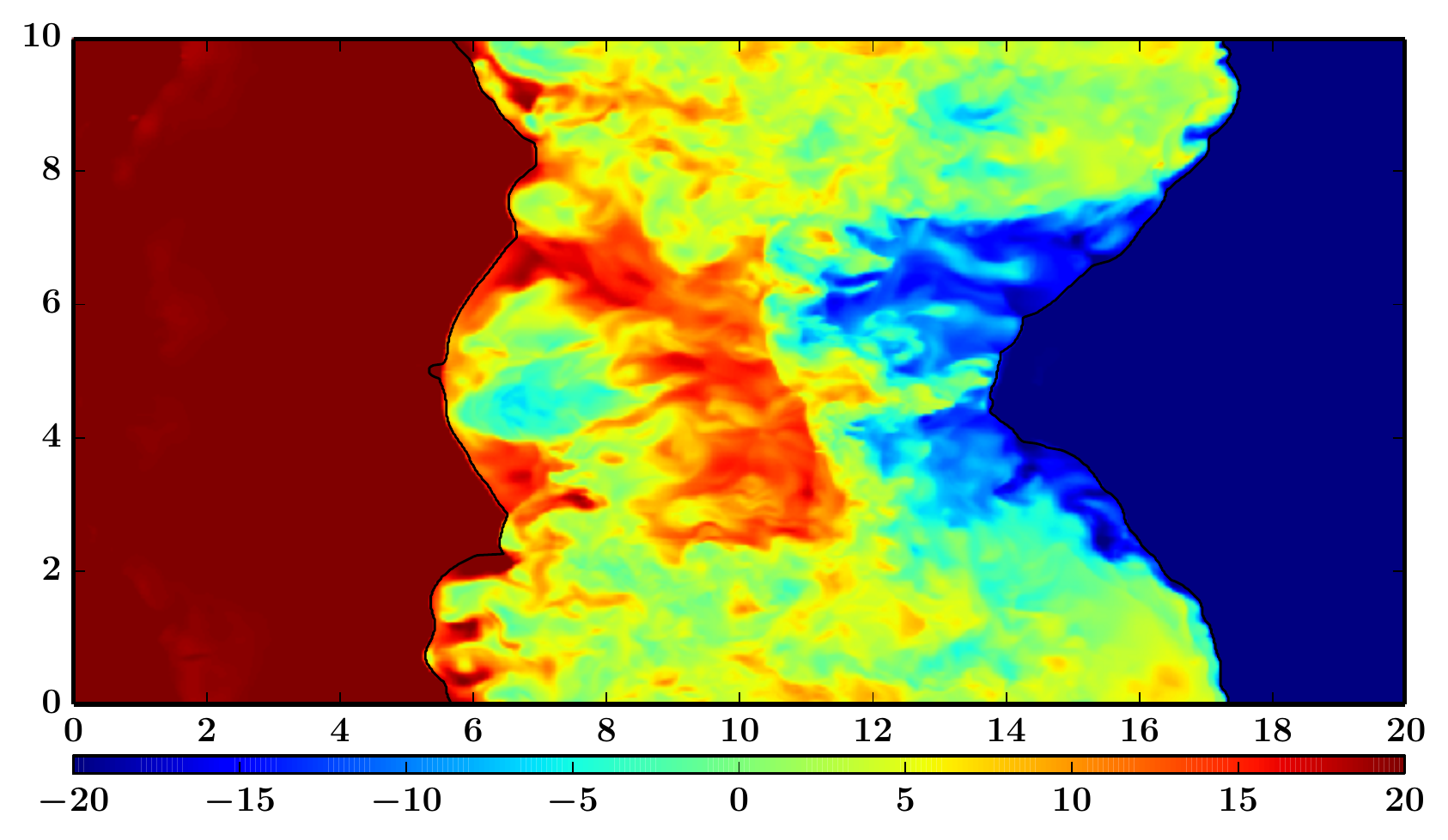}}
    \end{minipage}
    \hspace{0.5cm}
    \begin{minipage}[t]{0.47\textwidth}
        \centering{\includegraphics[scale=0.45]{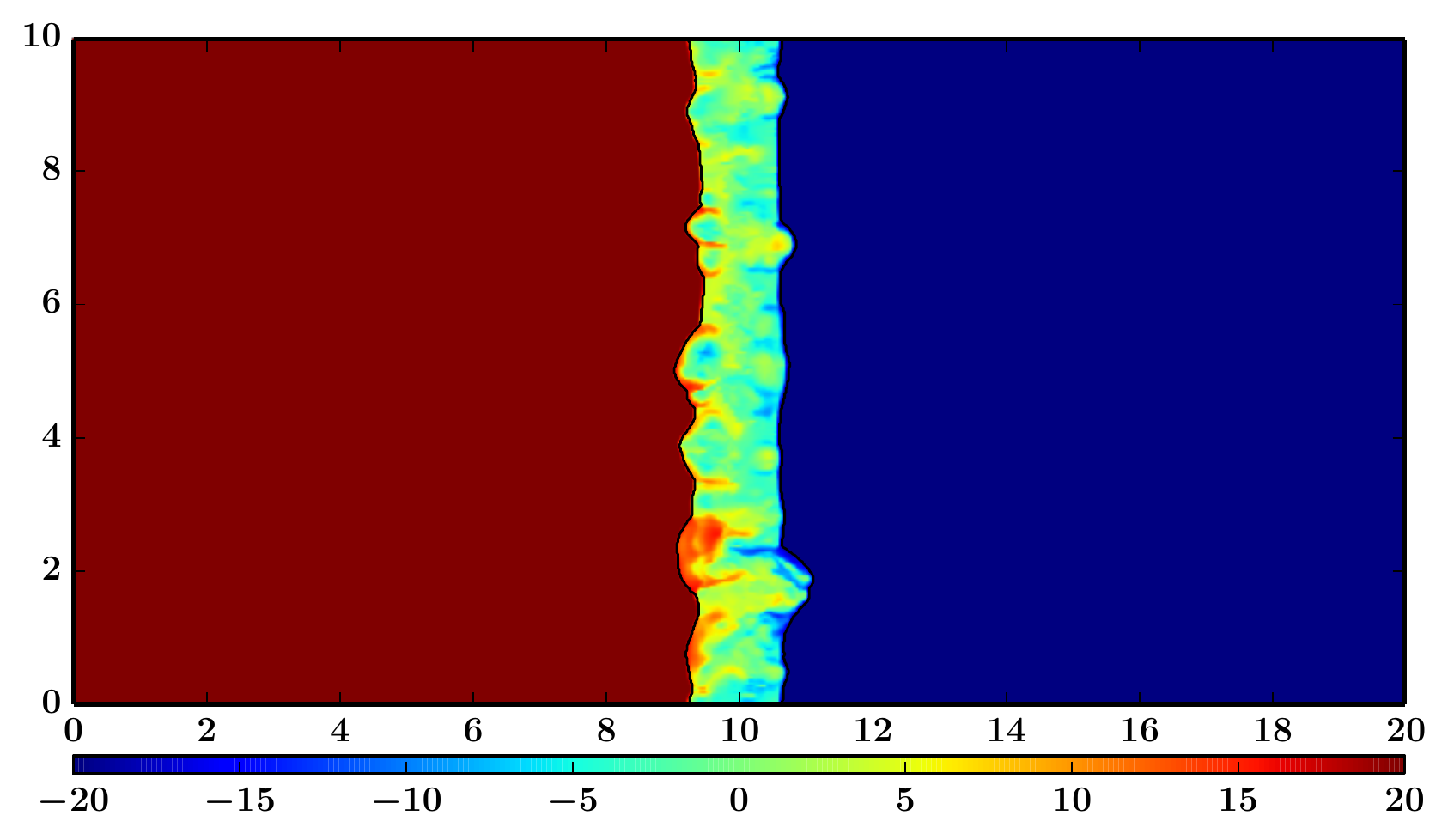}}
    \end{minipage}\\
    {\large (c) $\mydrhop =$ 100\% } 
    \hspace{6.0cm} {\large (d) $\mydrhop =$ 1\%}  \\
    \begin{minipage}[t]{0.47\textwidth}
        \centering{\includegraphics[scale=0.45]{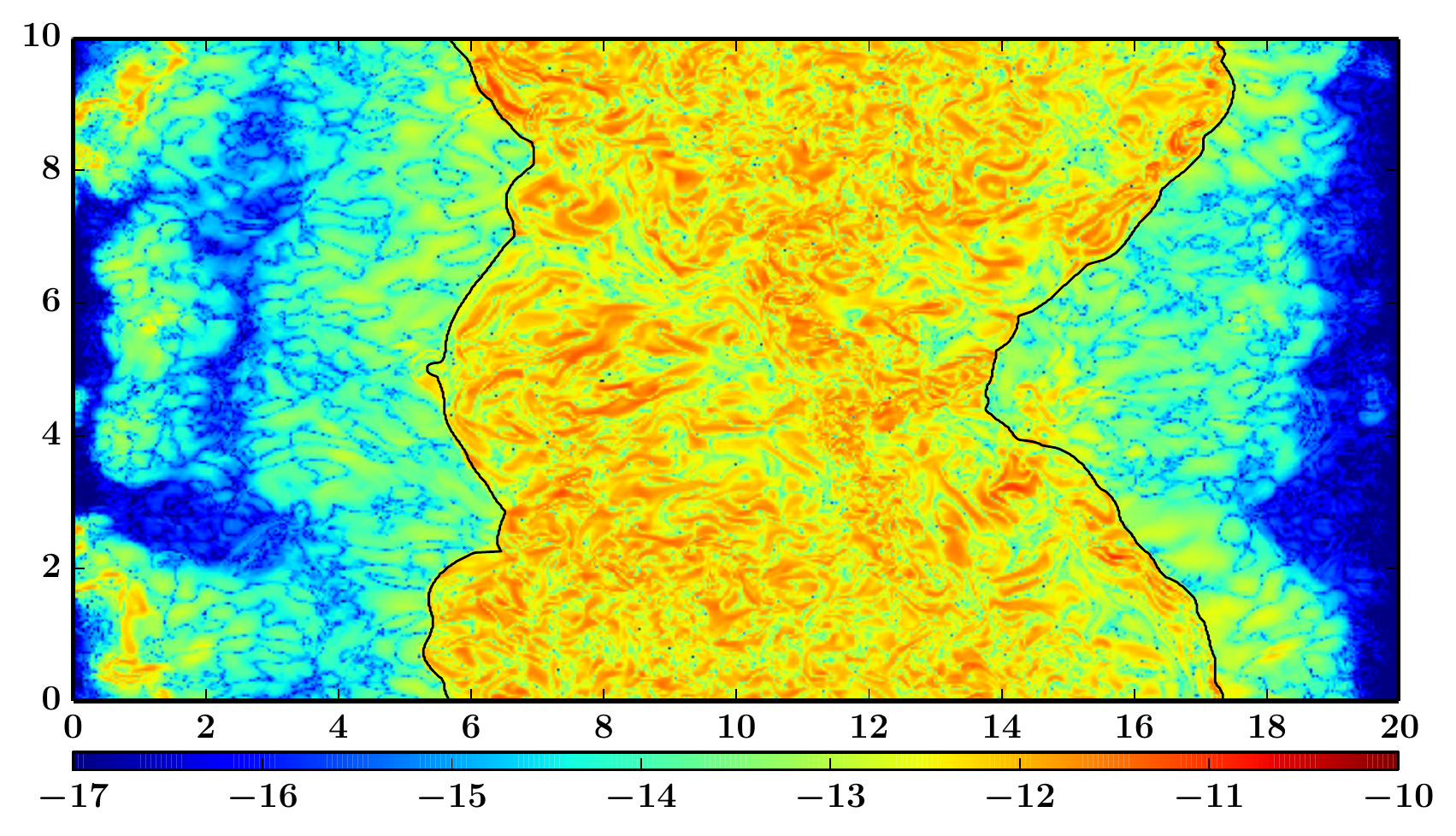}}
    \end{minipage}
    \hspace{0.5cm}
    \begin{minipage}[t]{0.47\textwidth}
        \centering{\includegraphics[scale=0.45]{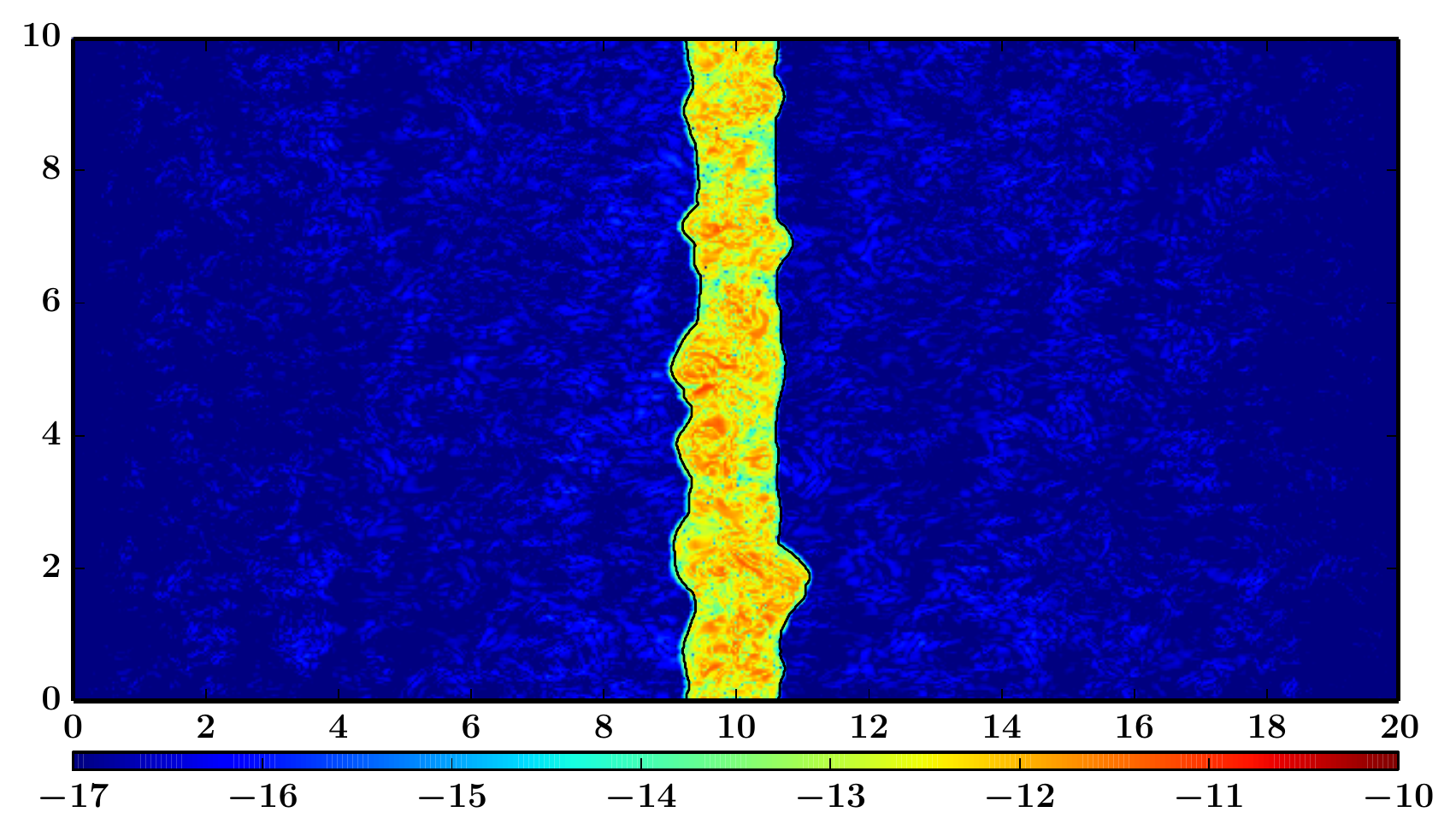}}
    \end{minipage}\\
    \vspace{-5pt}
    \caption{Panels (a) and (b): $v_x^{} [{\rm \kms}]$ slice on $z=5$ pc plane at $1.8$ Myr,
             where $\mydrhop=31.6$ \% (Run C2D0316, Panel (a)) and 
             $3.16$ \% (Run C2D0031, Panel (b)), both of which employ Phase $\myaa$ and $\Delta x =0.02$ pc.
             Panels (c) and (d): Same as Panels (a) and (b) but
             the $y$-component of vorticity, $\log \left( \left|(\nabla \times \mathbf{v})_{\rm y}\right| [{\rm s^{-1}}]\right)$.
             }
    \label{fig:f24f25f26f27}
\end{figure*}
\begin{figure}
    \centering{\includegraphics[scale=0.92]{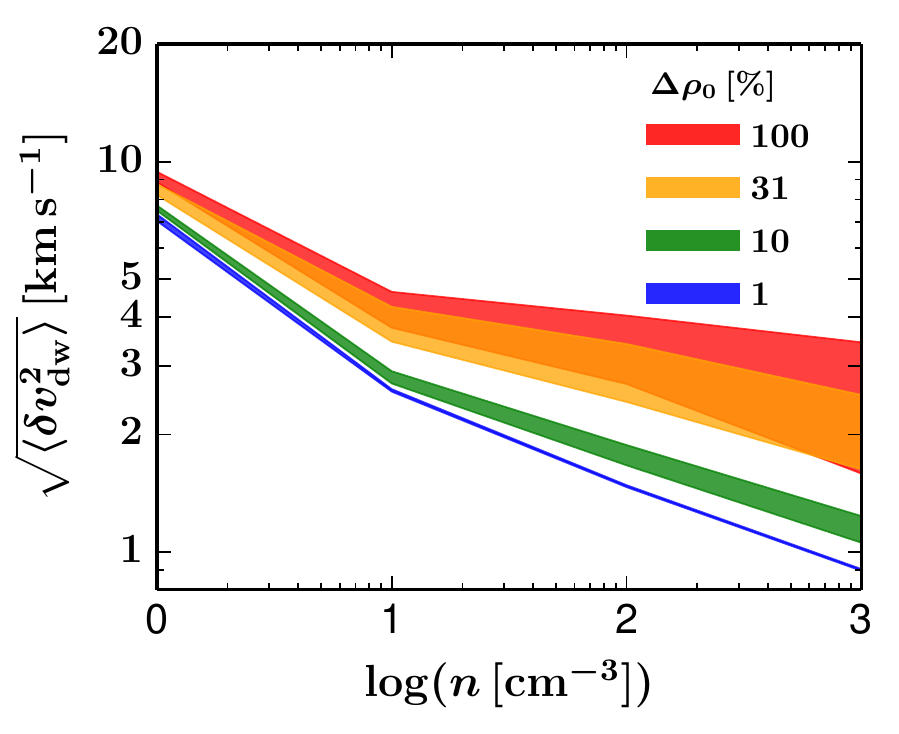}}
    \caption{The density-weighted velocity dispersion $\dvw$ as a function of $n$ and $\mydrhop$
             with $\Delta x= 0.02$ pc.
             The color corresponds to $\mydrhop$ and the shades show the range 
             of the maximum and minimum due to Phase $\myaa$, $\myab$, and $\myac$ in each $\mydrhop$.
             Note that the vertical axis is also in the logarithmic scale.
             Also note that the horizontal axis has logarithmically equal bins as $\log(n[\cmkk]) = [0,1), [1,2), 
             [2,3),[3,\infty)$ and each bin is plotted at their lower limit density.
             }
    \label{fig:f28}
\end{figure}
\begin{figure*}
    \begin{minipage}[t]{0.47\textwidth}
        \centering{\includegraphics[scale=0.95]{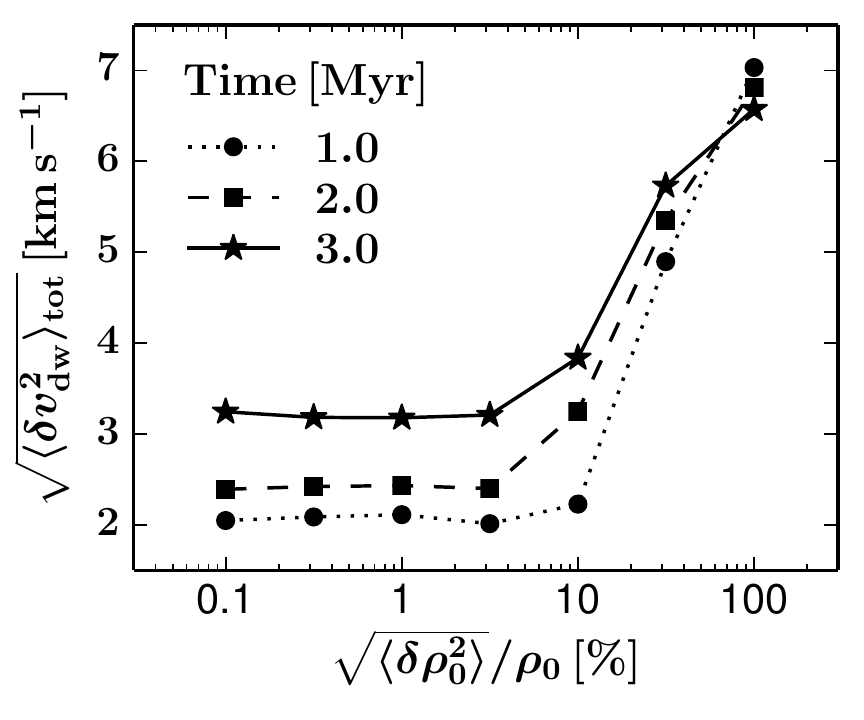}}
    \end{minipage}
    \hspace{0.5cm}
    \begin{minipage}[t]{0.47\textwidth}
        \centering{\includegraphics[scale=0.95]{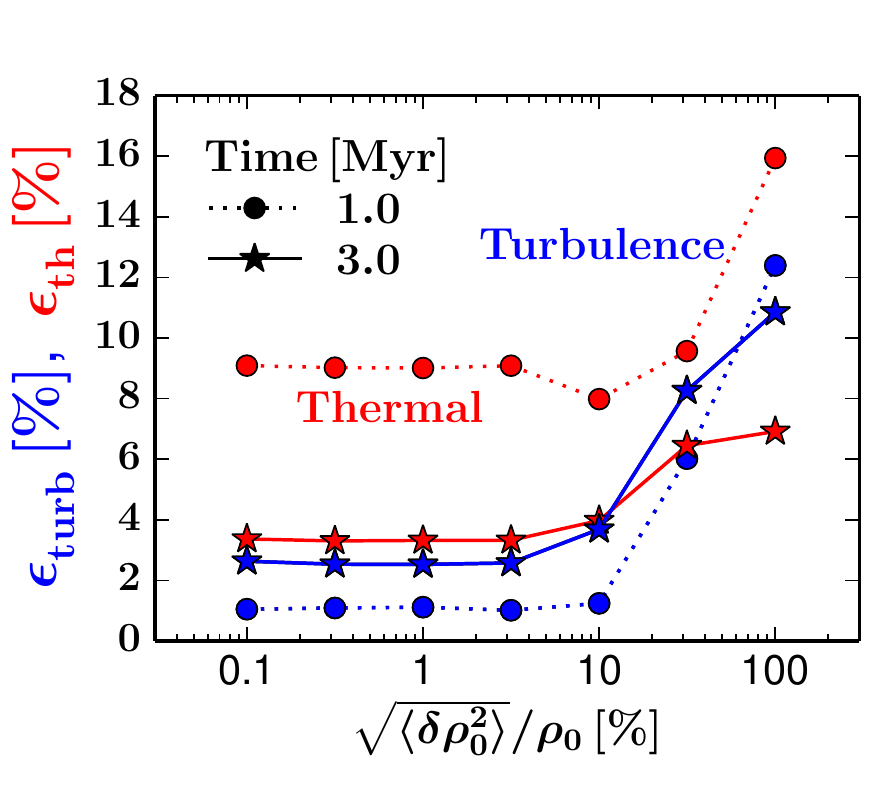}}
    \end{minipage}\\
    \caption{Left: the density-weighted velocity dispersion $\sqrt{\langle \delta v_{\rm dw}^2\rangle_{\rm tot}}$
             as a function of $\mydrhop$ at 1, 2, and 3 Myr. 
             The subscription ``tot'' means that we take the average over the entire volume of the shock-compressed layer
             and we take the average of three random phases at each time step.
             Right: the energy conversion rate from the upstream kinetic energy
             into the post-shock turbulence energy ($\epsilon_{\rm turb}^{}$, blue), 
             and into the post-shock thermal energy ($\epsilon_{\rm th}^{}$, red)
             as a function of $\mydrhop$ at 1 and 3 Myr.
    }
    \label{fig:f29f30}
\end{figure*}

\subsection{Geometry and Turbulence}
\label{subsec:myxshock}
We found that two distinct behaviors depending on $\mydrhop$ 
appear also on the geometry of the shock-compressed layer corresponding to large/small $\mydrhop$.
The left panel of Figure~\ref{fig:f22f23} summarizes the time evolution of the average shock front position $\myxshock$ 
for each
$\mydrhop$ with $\Delta x= 0.02$ pc. The shades in each $\mydrhop$ show the range 
of the maximum and minimum due to the three random phases $\alpha$.
This shows that the shock-compressed layer becomes wider (with larger variation
by different $\alpha$)
when $\mydrhop$ is larger, which 
one can visually confirm also
in the density snapshots 
already shown in Figure~\ref{fig:f3f4f5f6f7f8}.
In addition, we investigate 
the dispersion of $\myxshock$, defined as
\begin{equation}
    \sqrt{\langle \delta x_{\rm shock}^2 \rangle} = \frac{\sum_{y,z}^{} \sqrt{(x_{\rm R}^{}(y,z)  - \bar{x}_{\rm R})^2 + 
    (x_{\rm L}^{}(y,z)  - \bar{x}_{\rm L})^2 }}{2N_{yz}}\,,
\end{equation}
where $\bar{x}_{\rm R}$ and $\bar{x}_{\rm L}$ are the mean position of the right and left shocks, respectively,
and the results are summarized in the right panel of Figure~\ref{fig:f22f23}.
$\sqrt{\langle \delta x_{\rm shock}^2 \rangle}$ gradually grows in time in all $\mydrhop$ cases,
especially density inhomogeneity with larger $\mydrhop$ deforms shock fronts more significantly
whose amplitude reaches a fraction of 10 pc (the size of the cross section of the computational domain).
Given that the Kolmogorov spectrum in the upstream density fluctuation has
a larger power on larger scales (Equation~\ref{eq:rho_init}),
we expect that the shock deformation is apparent on larger scales. %with larger $\mydrhop$.
The shock front geometry of $\mydrhop=100$ \%, for example, indeed indicates the impact from
the largest-scale mode of $\lvert k \rvert = 2 \pi / L_{y}$ (\eg, Panel (a) of 
Figure~\ref{fig:f3f4f5f6f7f8}). Smaller $\mydrhop$ introduces less deformation
and the shock front geometry becomes closer to the completely straight front observed in $\mydrhop=0$ \%
(\eg, Panels (d) and (e) of Figure~\ref{fig:f3f4f5f6f7f8}).

The two modes in these geometries, combined with the WNM:CNM mass fraction
in Section~\ref{subsec:conv}, suggest that
large-scale oblique shocks induced by significant shock deformation with larger $\mydrhop$
suppresses the energy dissipation at the shock fronts, 
driving stronger turbulence and forming less CNM in the shock-compressed layer,
and vice versa for smaller $\mydrhop$ cases.
We thus investigate the velocity and vorticity structure of the shock-compressed layer 
and the results are shown
in Figure~\ref{fig:f24f25f26f27}. We here choose $\mydrhop=100$ \%  and $1$ \% 
to clearly highlight the difference between large and small $\mydrhop$.
There are indeed fast WNM flows that continue into the shock-compressed layer 
without significant vorticity generation when $\mydrhop$ is large
(\eg, $3$ pc $\lesssim y \lesssim 7$ pc in Panels (a) and (c)),
whereas most of the flow is well decelerated when $\mydrhop$ is small (in Panels (b) and (d)).
We also measure the density-weighted velocity dispersion $\dvw$
as a function of $n$ and $\mydrhop$, which is shown in Figure~\ref{fig:f28}.
The bimodal behavior depending on $\mydrhop$ appears also in this $\dvw$
as we expected;
the less-decelerated fast flow of the WNM in larger $\mydrhop$ cases
drive larger $\dvw$ with significant variation due to different $\alpha$.

Note that the high-density structures
predominantly contribute to
the turbulent energy density $\rho \dvws$ 
in any $\mydrhop$ cases
because the overall $n$-dependence of $\dvw$
is $\dvw \propto n^{-m}$ with $m\leq0.5$
in the range of $n=1$ -- $1000\,\cmkk$.
In addition, $\dvw$ is dominated by the $x$-component
because the converging-flow has a directionality in $x$.
We found that the $x$-component of $\dvw$ 
amounts to $>70$ \% of the total $\dvw$ in all density range
in any $\mydrhop$ cases.
Such anisotropic turbulence is reported even in magnetized converging flow simulations
when the magnetic fields are mostly parallel to the gas flow
\citep[\eg,][]{vazquezsemadeni2007,Inoue2012,Iwasaki2018}.

Also note that the overall decreasing trend of $\dvw$ with $n$ is consistent with other simulations and observations.
For example, the velocity dispersion $\geq 2\, \kms$ at $n<1\,\cmkk$
is reported from recent observations of the WNM absorption feature \citep[\eg,][]{Patra2018}.
The velocity dispersion $\leq2\,\kms$ at $n>10\,\cmkk$ is 
reported in previous converging WNM flow studies 
without magnetic fields \citep[\eg,][]{Heitsch2006b}.
\cite{Fukui2018} also report this decreasing trend of $\dvw$ with $n$, 
where they perform synthetic observations of
H{\sc i} line profiles in magnetohydrodynamics simulations
and compare with emission-absorption measurements along quasar line of sights\footnote[12]{\cite{Fukui2018}
report $\sim 5\, \kms$ for CNM and $\sim 40\, \kms$ for the WNM.
These values are a factor higher than 
our measurements. This may originate in the difference of inflow gas density:
\cite{Fukui2018} utilize magnetohydrodynamics converging flow simulations from 
\cite{Inoue2012} with $n=5.2\,\cmkk$ whereas we have $n=0.57\,\cmkk$.}.

\subsection{Energy Partition}
\label{subsec:eb}
Finally, we investigate the energy partition in the shock-compressed layer
and its dependence on $\mydrhop$.
The left panel of Figure~\ref{fig:f29f30} shows the time evolution of $\sqrt{\langle \delta v_{\rm dw}^2\rangle_{\rm tot}}$
as a function of $\mydrhop$. 
The subscription ``tot'' means that we take the average both over the entire volume of the shock-compressed layer
and over the three random phases.
This again clearly shows the bimodal behavior with respect to $\mydrhop$, as expected from Figure~\ref{fig:f28}.
In addition, we found that small $\mydrhop\leq 10$ \% creates the velocity dispersion of $2.0$ -- $3.2\,\kms$ 
even down to $\mydrhop = 0.1$ \%.
The value of $\sim 2\, \kms$ is consistent with the dispersion driven by the thermal instability alone \citep[\eg,][]{Koyama2002,Koyama2006}.
This result indicates that shocks are always able to drive the velocity dispersion of $\sim 2\,\kms$
through the thermal instability even when they propagate though the ISM with an almost uniform density.

Based on this $\sqrt{\langle \delta v_{\rm dw}^2\rangle_{\rm tot}}$, we measure
the conversion rate of the upstream kinetic energy into the post-shock turbulent and thermal energy as
\begin{align}
    &\epsilon_{\rm turb}^{} (t) = \frac{M_{\rm total}(t)\, \langle \delta v_{\rm dw}^2(t)\rangle_{\rm tot}/2}{\int \dot{M}_{\rm total}(t)\,\myvin^2 /2\,{\rm d}t}
    = \frac{\langle \delta v_{\rm dw}^2(t)\rangle_{\rm tot}}{\myvin^2} \,, \label{eq:teps1} \\
    &\epsilon_{\rm th}^{} (t) = \frac{\int P(\mathbf{x},t) {\rm d}\mathbf{x} /(\gamma -1)}{\int \dot{M}_{\rm total}(t)\,\myvin^2 /2\,{\rm d}t} 
    = \frac{\langle P(t)\rangle_{\rm tot}/(\gamma -1)}{\langle \rho(t) \rangle_{\rm tot} \myvin^2 / 2} \,. \label{eq:teps2}
\end{align}
Here $\dot{M}_{\rm total}(t)$ denotes the mass accretion rate into the shock-compressed layer at time $t$,
which can be approximated as $\dot{M}_{\rm total}(t) \simeq 2 \rho_0 L_y L_z (\myvin+\myvshock(t))$.
Note that $\myvin = 20\,\kms = {\rm Const.}$ and we can assume 
$\myvshock = {\rm Const.}$ as a zeroth-order estimation based on the time-evolution of $\myxshock$.
Therefore, we use $M_{\rm total}(t) = \int \dot{M}_{\rm total}(t) {\rm d} t$ 
to obtain the second equality both in Equations~\ref{eq:teps1} and~\ref{eq:teps2}\footnote[13]{We
focus on the kinetic energy alone in the denominators because it dominates the inflow energy budget.
The denominators change by a factor $\sim 1.5$ when combined with the thermal energy
as $V_{\rm in}^2/2 + C_{\rm s}^2/(\gamma-1)$.}.

The right panel of Figure~\ref{fig:f29f30} shows $\epsilon_{\rm turb}^{}$ and $\epsilon_{\rm th}^{}$
as a function of $\mydrhop$ and time.
The bimodal behavior depending on $\mydrhop$
again appears in $\epsilon_{\rm turb}^{}$ and $\epsilon_{\rm th}^{}$ individually,
and appears also in the relative importance of $\epsilon_{\rm turb}^{}$ and $\epsilon_{\rm th}^{}$;
for example, at 3 Myr,
the turbulent and thermal pressures equally support the shock-compressed layer when $\mydrhop \leq 10$ \%
whereas the turbulence dominates when $\mydrhop > 10$ \%.

In small $\mydrhop$ cases, the shock-compressed layer is initially supported by the thermal pressure
with limited turbulence. As CNM formation proceeds and the turbulence is developed (as seen in 
the left panel of Figure~\ref{fig:f29f30}), the turbulent and thermal pressures start to equally support
    the shock-compressed layer. The sum of $\epsilon_{\rm turb}^{}$ and $\epsilon_{\rm th}^{}$ is limited to 
$<10$ \% of the injected kinetic energy due to
the radiation energy loss (implemented as the source term $\rho \myCooL$ in Equation~\ref{eq:EE}).
In larger $\mydrhop$ cases,
strong turbulence prevents the dynamical condensation by cooling
and the following CNM formation.
The shock-compressed layer is occupied by the low-density high-temperature WNM
and UNM
(see the $P-n$ diagram of Figure~\ref{fig:f10f11f12f13}),
which keeps $\epsilon_{\rm th}^{}$
higher than that in smaller $\mydrhop$ cases.
The turbulence dominantly supports the shock-compressed layer at 3 Myr
and $\epsilon_{\rm turb}^{}$ reaches $12.4$ \% at maximum. 
Such a high conversion rate is consistent with the one driven by the interaction 
between shocks and density inhomogeneity 
\citep[\eg,][]{Inoue2012c,Inoue2013b,Iwasaki2018}.
\citealt{Inoue2013b} showed that the growth velocity of the Richtmyer-Meshkov instability
\citep{Richtmyer1960,Nishihara2010} is able to account for the velocity dispersion of the turbulence.

These results indicate that the observed supersonic turbulence 
within molecular clouds are originated not only from a weak turbulence of a few $\kms$ 
by the thermal instability, but also (or even overridden) by 
the interaction between shocks and density inhomogeneity
because 
the ISM in reality 
have the density fluctuation
close to $\mydrhop = 100$ \% (see also Section~\ref{subsec:mf}).
This is also consistent with the conclusion of 
\cite{Inoue2012}.

\section{Discussions}
\label{sec:discussions}
\subsection{Mass Fraction}
\label{subsec:mf}
As already seen in Figure~\ref{fig:f18f19f20f21} and Table~\ref{table:resol},
the CNM mass fraction varies from $45$ \% to $70$ \% as $\mydrhop=0.1$ \% to $100$ \% at 3 Myr.
%the WNM:CNM mass fraction varies from $1:9$ to $4:6$ as $\mydrhop=0.1$ \% to $100$ \%
%at 3 Myr.
There is also a time evolution such that $f_{\rm CNM}$ is initially 
0 because we inject the WNM flows, and gradually increase to the aforementioned values.
Typically after the typical cooling time of the injected WNM flow of $\mytaucool =1.3$ Myr, 
$\myxshock$ evolves almost linearly with time
where the CNM mass fraction is constant (\ie, the mass conversion rate into the WNM and CNM
per injected WNM is constant, and therefore the layer widens quasi-steadily).
The time-evolution of the mean density $\langle n \rangle$ reflects such evolution
as shown in Figure~\ref{fig:f14f15f16f17}.

We suggest that the mass fraction
%ratio 
measured at 3 Myr is a typical value achieved in the ISM\@.
Any volume of the ISM is typically swept up by at least one supersonic shock
in every 1 Myr \citep[\cf,][]{McKee1977} due to supernovae.
Thus, once the ISM reaches the quasi-steady state as observed at 3 Myr in our simulations, 
such frequent shock events presumably sustain that state.
We also suggest that 
the density structure with $\mydrhop = 100$ \% is more common in the ISM
because the density probability distribution function in solar neighborhood star-forming regions 
has a wide density range well fitted by a log-normal function (\eg, measured from dust: \citealt[]{Schneider2013,Schneider2016}).
Therefore, our results suggest that the typical CNM mass fraction is $\sim 50$ \% in the typical ISM.
Large-scale simulations overestimate dense gas mass available for star formation
if their sub-grid models immediately convert 100 percent of the diffuse WNM into CNM and molecular gas within 3 Myr.
It is left for future studies to numerically simulate another shock passage 
through the multiphase ISM 
whose CNM mass fraction is already $\sim 50$ \%
%whose WNM\mk{+UNM}:CNM mass ratio is already 1:1
\citep[\cf,][]{Inoue2012,Iwasaki2018}.

\subsection{Variation due to the Non-linear Nature}
\label{subsec:sto}
The CNM formation in any converging flow simulation 
has a chaotic behavior due to the non-linear nature of the system,
in a sense that a slight difference in the shock deformation
later results in the formation of CNM clumps with slightly different masses and sizes,
and this impacts subsequent CNM formation by altering the density/turbulent structures.
For example, as shown in Panels (d) and (e) of Figure~\ref{fig:f3f4f5f6f7f8},
even a limited amplitude difference between $\mydrhop=3.16$ \% and $1$ \% 
resulted in a different density distribution in the shock-compressed layer already by $1.8$ Myr.
Such a chaotic behavior
always occurs also within a given $\mydrhop$ between
various $\alpha$ under a fixed $\Delta x$ and between various $\Delta x$
under a fixed $\alpha$, which is able to introduce variation in the mean properties of the layer.
Our results in Sections~\ref{subsec:conv},~\ref{subsec:myxshock}, and~\ref{subsec:eb} show
that, when $\mydrhop$ is small, 
this non-linear nature
is less prominent 
if we average over the shock-compressed layer
as $\myxshock$ and $\langle n \rangle$.
This is because $>90$ \% of the WNM flow kinetic energy is dissipated and most of the injected mass condense into CNM
with limited turbulence.
Therefore, the monotonic convergence with $\Delta x$ in the mean properties 
directly reflects whether or not we resolve the condensation process well enough.

In contrast when $\mydrhop$ is large, such a chaotic behavior
appears in the mean properties,
which we found is primarily due to the interaction between shocks and CNM clumps.
In a large $\mydrhop$ case, the layer becomes strongly turbulent 
where CNM clumps have a large velocity dispersion;
for example, $\dvw$ at $n=100$ -- $1000\, \cmkk$ reaches 
$2.7$ -- $4.0\, \kms$ when $\mydrhop=100$ \%,
which is $>5$ times faster than the CNM sound speed ($0.67\, \kms$ 
at the thermally balanced state of 
$n=100\, \cmkk$ and $T=41\,{\rm K}$).
Once those fast CNM clumps form, 
they sometimes push/penetrate shock fronts (\eg, 
as seen at $y\simeq1$ pc and $5$ pc in Panel (b) 
and $y\simeq1$ pc in Panel (c) of Figure~\ref{fig:f3f4f5f6f7f8}).
This process provides an additional deformation of shock fronts on top of the deformation
driven by the upstream density inhomogeneity.
This additional deformation 
impacts the following
CNM formation and turbulence in the shock-compressed layer,
which introduce another shock deformation subsequently.
The mean properties (\eg, the CNM mass fraction) also vary accordingly.
This process
always prevents a perfect convergence,
and 
our results in Sections~\ref{subsec:conv},~\ref{subsec:myxshock}, and~\ref{subsec:eb} suggest that
such variation 
due to different $\Delta x$ becomes comparable 
to the variation due to different $\alpha$
when $\mydrhop$ is large.

\subsection{Self Gravity}
\label{subsec:sg}
We by now ignore self-gravity in this article.
Self-gravity decelerates the expansion of the shock-compressed layer,
keeps CNM clumps stay in the layer,
and pulls CNM clumps back to the layer
even when they penetrate the shock fronts.
Let us estimate two timescales determined by self-gravity
and show the time range where
the absence of self-gravity is valid.

Firstly, we define a timescale after which the shock-compressed layer
becomes self-gravitating rather than the ram pressure confined.
Let us label this timescale as $\mytsgl$.
$\mytsgl$ can be estimated as the force balance
between the self-gravity and the ram pressure of the converging flows
as $\pi G \Sigma^2(t) /2 > \rho_0 V_{{\rm in}}^2$,
where $G$ is the gravitational constant, and $\Sigma(t)$ is the 
column density of the shock-compressed layer at time $t$.
Given that 
$\Sigma (t) \simeq 2 \rho_0 \myvin t$, %in a converging flow system,
$\mytsgl$ can be estimated as 
\begin{equation}
    \mytsgl \simeq \sqrt{\frac{1}{2 \pi G \rho_0}} \, = \, 44 \, \mathrm{Myr} \left( \frac{n_0}{0.57 \, \cmkk} \right)^{-1/2}\,.
\end{equation}

Secondly, 
there is a typical timescale over which the self-gravity 
of the shock-compressed layer pulls back 
CNM clumps that push/penetrate the shock fronts \citep[see Appendix~C in][]{Iwasaki2018}.
Let us label this timescale as $\mytstop$,
which is $\mytstop \simeq \left(\myvej/4 \pi G \rho_0 V_{\rm in} \right)^{1/2}$
where $\myvej$ is the CNM clump's ejection velocity when they penetrate the shock fronts.
$\mytstop$ can be estimated as
\begin{equation}
    \mytstop \simeq 12 \, \mathrm{Myr} \, \left(\frac{\myvej}{3 \, \kms}\right)^{\frac{1}{2}} 
    \left(\frac{n_0}{0.57 \, \cmkk}\right)^{-\frac{1}{2}}
    \left(\frac{\myvin}{20 \, \kms}\right)^{-\frac{1}{2}} \,.
\end{equation}
Here we take $3\, \kms$ for $\myvej$ based on the typical CNM $\dvw$ 
in our simulations with $\mydrhop=100$ \%
(see Figure~\ref{fig:f28}).

Therefore both timescales suggest that it is an acceptable assumption to ignore self-gravity,
when we focus on the early stages of the multiphase ISM formation
as we did in our simulation $\lesssim 3$ Myr,

\subsection{Role of Thermal Conduction}
\label{subsec:FL_TC}
The thermal conduction plays an important role in determining the detailed structure of the CNM, 
especially the structure of a thin transition layer between the WNM and CNM \citep{Field1965}. 
This is characterized by the Field length of the CNM\@.
Under the cooling function and conduction rate (Equations~\ref{eq:EE} and~\ref{eq:hcfunc}),
the typical Field length of the CNM is as short as $10^{-3}$ to $10^{-4}$ pc (at T=20 K and n=150 cm$^{-3}$;
see \cite{Koyama2004}),
which we do not resolve in our current simulations.
However, the thermal conduction does not appear to have a strong influence on 
the convergence in the macroscopic properties of the shock-compressed layer (Section~\ref{subsec:conv}).
This is because the dynamics in our simulation is dominated more by 
the interaction between the shocks and upstream density inhomogeneity,
and also by the dynamical condensation due to cooling,
than by the thermal conduction alone.
We describe our understanding on this situation below.

In a system where the thermal conduction controls the dynamics, 
it is important to resolve the Field length.
For example, starting from a thermal equilibrium UNM, 
\cite{Koyama2004} performed a one-dimensional numerical calculation 
to investigate the formation of the WNM and CNM, 
and follow the long-term evolution of the motion between the two phases.
They showed that the thermal conduction drives the motion of $\sim 0.1\, \kms$,
and it is required to resolve the Field length to calculate this motion 
(by at least three cells: Field condition).
\cite{Iwasaki2014} investigated the two-dimensional cases, 
which also observed the $\sim 0.1\, \kms$ velocity dispersion driven by the thermal conduction
and confirmed the requirement of the Field condition 
to achieve the convergence in the velocity dispersion.

In contrast, in more dynamical systems like our simulations, 
super-sonic shocks create the WNM and UNM with high pressure, which is
far from the thermal equilibrium. In this case, 
the cooling dominates the dynamical evolution of those phases 
\citep[see \eg, Appendix A of][]{Iwasaki2012},
and it is important to resolve the cooling length.
\cite{Koyama2002} numerically investigated the evolution of the post-shock medium and showed that 
the interaction between the shocks and upstream density inhomogeneity, 
as well as the following dynamical condensation of the WNM and UNM into CNM 
due to cooling, keep driving much faster turbulence of a few $\kms$.
\cite{Hennebelle2007a} numerically investigated the effect of the thermal conduction
in their high resolution converging-flow simulations 
(with 0.002 pc resolution albeit two dimensional) by changing the thermal conductivity,
and they have shown that the total mass in the CNM does not significantly change 
with the thermal conductivity.

Therefore, the thermal conduction does not play a critical role in our converging-flow calculations, 
and this seems to be the reason why we do not necessarily resolve the Field length of $10^{-3}$ pc in this case.
%and it is thus not a stringent requirement to resolve the Field length of 10^-3 pc. 
Nevertheless, it is still important to resolve the cooling length
of a few pc -- $10^{-2}$ pc to achieve the convergence in the macroscopic properties 
of the shock-compressed layer
(as shown in Panels (c) and (d) of Figure~\ref{fig:f10f11f12f13} and discussions therein).

\subsection{Effective EoS with $\mygamma$ and its Application}
\label{subsec:g_eff}
ISM models and star formation prescription below the spatial resolution
is one of the challenges and uncertainties in large-scale simulations (\eg, 
evolution of the entire galactic disk and large-scale structure of the Universe).
To consistently calculate such sub-resolution scale ISM evolution and star formation,
there have been semi-analytical studies aiming at formulation
of an ISM effective equation of state (EoS) that 
describes the balance between the CNM formation by the thermal instability
and supernovae feedback \citep[\eg,][]{Yepes1997,Springel2003}.
There are also studies based on numerical simulations
to model such 
ISM effective EoS controlled by turbulence \citep{Joung2009,Birnboim2015}.
In this section, we would like to propose 
a similar effective equation of state
on a $\sim 10$ pc scale
that 
approximates the multiphase ISM in 
the CNM formation epoch as a one-phase medium.
This is
in a form of $P\propto  \rho^{\mygamma}$,
where we evaluate the effective index $\mygamma$ based on the results of our converging-flow simulations.

\begin{figure}\centering{
\includegraphics[width=0.95\columnwidth,keepaspectratio]{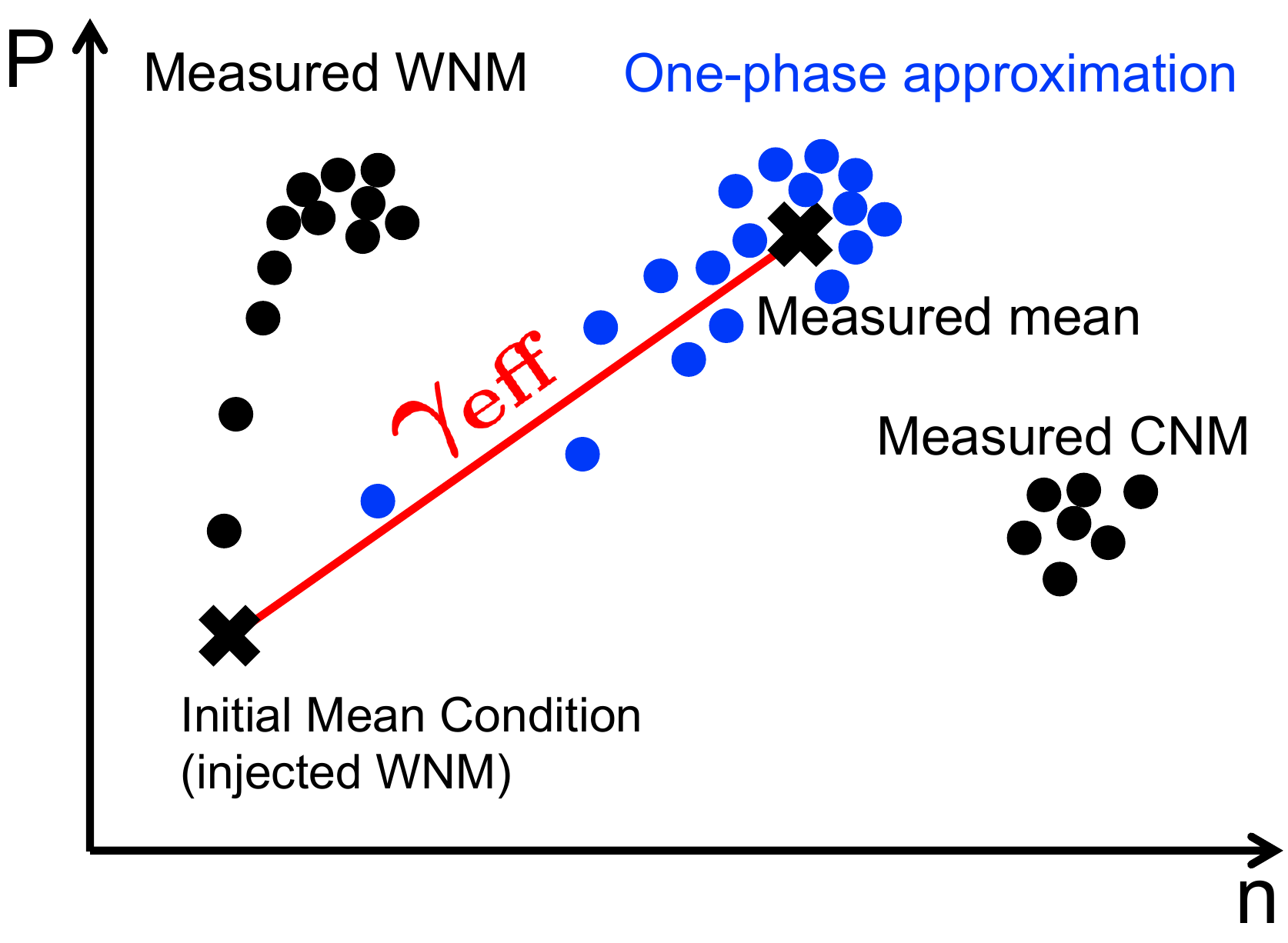}}
\caption{Schematic $P$-$n$ diagram to explain 
    the concept of $\mygamma$ formulated from our converging-flow simulations.
    The density of filled circles schematically represents the mass frequency 
    in the shock-compressed layer (see Panels (a) and (b)
    of Figure~\ref{fig:f10f11f12f13}).
    The bottom-left black cross shows the injected WNM state.
    In converging-flow simulations,
    the injected WNM is first adiabatically shocked (``Measured WNM'', black circles on the left)
    then cools to form CNM (``Measured CNM'', black circles on the right).
    As a result, we obtain the mean state averaged over the shock-compressed layer 
    (``Measured mean'', the top-right black cross).
    $\mygamma$ connects the initial WNM state and the mean state of the multiphase ISM 
    (``$\mygamma$'', the red solid line).
    The converging flow of an adiabatic WNM using the effective index $\mygamma$ 
    reproduces the mean state of the multiphase ISM
    without directly calculating the heating and cooling processes (``one-phase approximation'', blue circles).
    }
\label{fig:f31}
\end{figure}

The concept of our effective EoS is summarized in Figure~\ref{fig:f31}.
The ordinary Rankine-Hugoniot relations connect
physical quantities across a shock front 
in a one-dimensional adiabatic flow.
The density ratio of the post/pre shock regions, $r=\rho_2/\rho_1$, 
is characterized as
$r = (\gamma+1)\MachN^2/((\gamma-1)\MachN^2+2)$,
where $\MachN$ is the Mach number $\MachN = \myvinp/\mycs$ 
with $\myvinp$ as the flow speed in the shock-front rest-frame,
and $\gamma$ represents the polytropic index of the fluid.
This relation can be inverted as
$\gamma = (2r -\MachN^2 (1+r))/(\MachN^2 (1-r))$
to evaluate $\gamma$ 
by measuring density ratio $r$ and $\MachN$ of that fluid.
As an analogy from such adiabatic shocks, 
we propose that measurement of $r$ and $\MachN$ of the multiphase ISM
should also give an effective index $\mygamma$,
which 
approximates the multiphase ISM as a one-phase medium.
Here, the one-phase approximation
means that 
adiabatic WNM with the EoS using $\mygamma$ evolves 
while having its mean properties consistent with that of the multiphase ISM 
even without directly solving heating and cooling processes.
Such an effective EoS based on our converging-flow simulations
should be a relation 
of $P \propto \rho^{\mygamma}$
that connects the initial state of the injected WNM
and the simulated mean state of the shock-compressed layer (see Figure~\ref{fig:f31}).
Qualitatively speaking, when the mean density of the layer is low 
and the layer is geometrically widen,
the multiphase ISM is {\it stiff}\ against a given ram pressure by the WNM inflow,
and the corresponding $\mygamma$ is expected to be large.
When the mean density is high and the layer is geometrically narrow,
the multiphase ISM is {\it soft}\
and the corresponding $\mygamma$ is expected to be small.

The converging flow configuration is almost in the post-shock 
rest-frame whereas shock propagations in reality are mostly in the pre-shock rest-frame.
Therefore, we need to modify the original Rankine-Hugoniot relation as follows
when evaluating $\mygamma$ from our simulations: 
\begin{equation}
    \mygamma = \frac{2\myrj -\MachNs^2 (1+\myrj)}{\MachNs^2 (1-\myrj)} \,.
    \label{eq:g_eff_def}
\end{equation}
Here $\myrj$ is the effective density ratio, $\myrj=\langle \rho \rangle/\rho_0$,
where $\langle \rho \rangle$ is the mean mass density in the shock-compressed layer.
This $\langle \rho \rangle$ includes both the WNM and CNM because 
we aim at formulating an EoS representing the overall mean properties of the multiphase ISM.
$\MachNs$ is the Mach number in the shock-front rest-frame and therefore
$\MachNs = (\myvin+\myvshock)/\mycswnm$
where $\myvshock$ is the shock propagation speed. %in converging systems. %(\ie, in the post-shock rest frame).

\begin{figure*}
    \begin{minipage}[t]{0.47\textwidth}
        \centering{\includegraphics[scale=0.9]{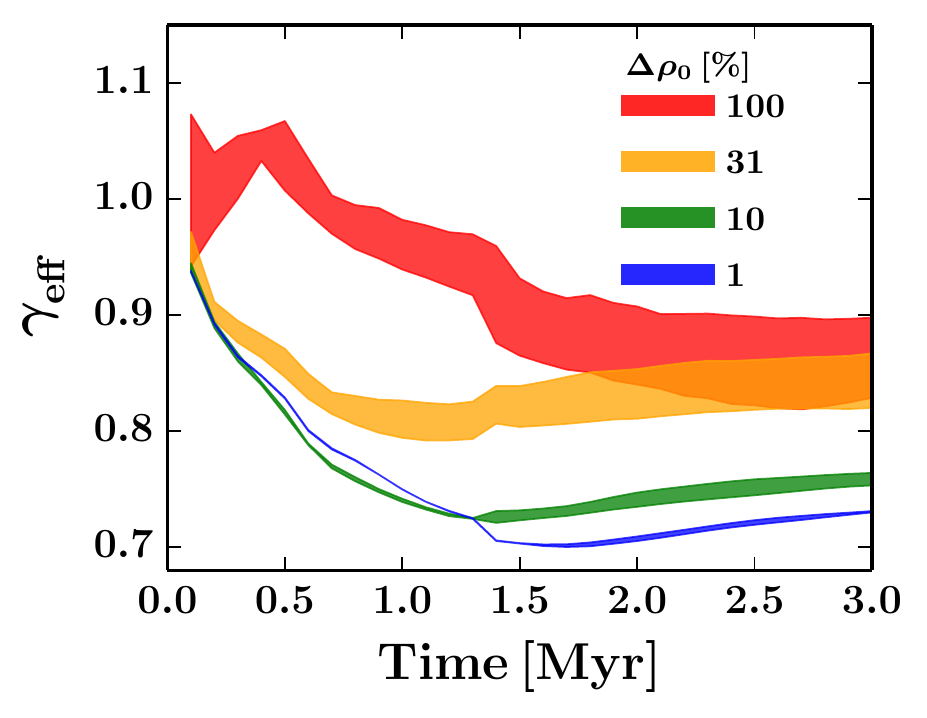}}
    \end{minipage}
    \hspace{0.5cm}
    \begin{minipage}[t]{0.47\textwidth}
        \centering{\includegraphics[scale=0.9]{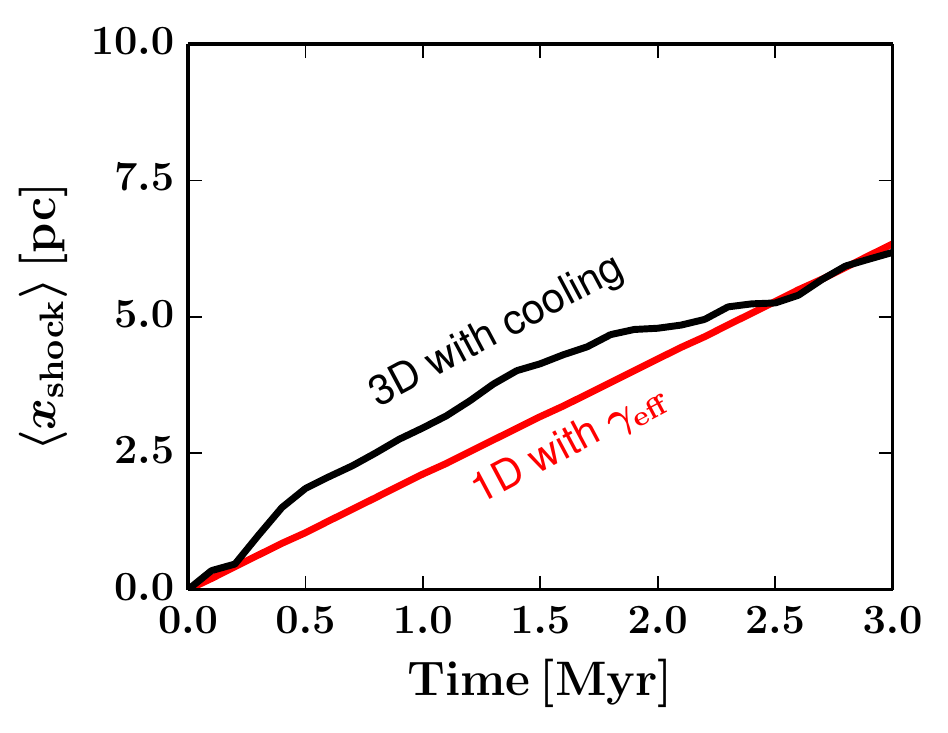}}
    \end{minipage}\\
    \caption{Left: The time evolution of the effective index $\mygamma$ as a function of time and $\mydrhop$
             with $\Delta x= 0.02$ pc.
             The color corresponds to $\mydrhop$ and the shades show the range 
             of the maximum and minimum due to Phase $\myaa$, $\myab$, and $\myac$ in each $\mydrhop$.
             Right: the time evolution of the average shock front position $\myxshock$.
             ``3D with cooling'' (black line) shows $\myxshock$ from $\mydrhop=100$ \% (Run C2D1000 with Phase $\myaa$),
             whereas ``1D with $\mygamma$'' (red straight line) shows $\myxshock$ of the test demonstration
             using $\mygamma=0.863$ (Run E2D0000).
             }
    \label{fig:f32f33}
\end{figure*}

The left panel of Figure~\ref{fig:f32f33}
shows the time evolution of  $\mygamma$.
We measure the $\myvshock$ as $(x_{\rm shock}(t)-x_{\rm shock}(t-0.1\,\mathrm{Myr}))/0.1\,\mathrm{Myr}$ 
until $\mytaucool$ (=1.3 Myr), 
and perform a linear fit as $(x_{\rm shock}(t)-x_{\rm shock}(\mytaucool))/(t-\mytaucool)$ after $\mytaucool$.
All $\mydrhop$ cases show a decreasing trend of $\mygamma$ in time,
which reflects the time-evolution of $\langle n \rangle$ due to 
the CNM formation (\ie, the layer becomes softer by forming CNM;
see Figure~\ref{fig:f14f15f16f17}).
$\mygamma$ tends to become constant after $\mytaucool$ because 
the layer expansion becomes quasi-steady with an almost quasi-steady NM mass fraction
(see the left panel of Figure~\ref{fig:f22f23}).
Larger/smaller $\mydrhop$ cases have lower/higher mean density,
and $\mygamma$ is stiffer/softer accordingly, as we expected.

We found that $\mygamma$ is softer than isothermal even in the stiffest case $\mydrhop =100$ \%,
and that it becomes as soft as $\mygamma \sim 0.7$ in smaller $\mydrhop$ conditions.
Such a high compressibility indicates that
the overall dynamics on $<10$ pc scales in large-scale simulations could be further improved 
by introducing this type of the effective EoS, especially 
in regions where the multiphase ISM formation is ongoing
without any feedback,
because most of current simulations use $\mygamma>1$ as a sub-grid model (\cf, \citealt{Inoue2019}).

As the first step towards such an actual application, 
we perform a converging adiabatic WNM flow by employing $\mygamma$
to demonstrate how well our $\mygamma$ reproduces the properties of the multiphase ISM.
In this demonstration, 
we directly update the thermal pressure based on the density evolution and $\mygamma$ at each timestep,
instead of calculating the heating and cooling ($\rho \myCooL$ in Equation~\ref{eq:EE}).
For simplicity, 
we opt to solve equations only in the $x$ direction to set this demonstration of $\mygamma$ in one-dimensional case,
and set a constant value of $\mygamma=0.863$
without modelling any time-evolution seen in the left panel of Figure~\ref{fig:f32f33}.
This value $0.863$ is based on the result of Run C2D1000 with Phase $\myaa$ at 3 Myr.
We also employ $\mydrhop=0$ to simulate a completely uniform head-on collision,
just as an simple demonstration.

The right panel of Figure~\ref{fig:f32f33} shows the time evolution of $\myxshock$.
``3D with cooling'' shows the result of our simulations with $\mydrhop=100$ \% (Run C2D1000 with Phase $\myaa$),
whereas ``1D with $\mygamma$'' shows the result of this the test demonstration using $\mygamma=0.863$ (Run E2D0000).
Table~\ref{table:geff} summarizes the measured properties at 3 Myr.
Our constant $\mygamma$ 
does not reproduce the detailed time-evolution of ``3D with cooling.''
Nevertheless it successfully reproduces 
$\myvin+\myvshock$, $\myxshock$, and $\langle n \rangle$ 
at 3 Myr within 11 \% difference,
which is still 
smaller than the variation due to different $\alpha$;
for example, 
the variation of $\myxshock$ due to Phases $\myaa$, $\myab$, and $\myac$ at 3 Myr
is $1.763$ pc in Run C2D1000 ($\sim$ 29 \% variation against $\myxshock=6.183$ pc
with Phase $\myaa$;
see the shade of $\mydrhop=100$ \% in the left panel of Figure~\ref{fig:f22f23}).

Ideally, we 
would like to provide a time-evolving model of $\mygamma (t)$
along with a time-evolving model of $f_{\rm CNM}$,
which, however, we reserve for future studies at this moment.
In such studies, we should also consider the intrinsic variation
of $\mygamma$ due to random phases, for example,
$\mygamma = 0.829$, $0.863$, and $0.898$
in Run C2D1000 with Phases $\myac$, $\myaa$, and $\myab$ at 3 Myr (see the left panel of 
Figure~\ref{fig:f32f33}).

\begin{table}
    \caption{Measured Properties at 3 Myr}
    \centering{
        \begin{tabular}{c|ccc}
            \hline
            \hline
            Runs                          & $\myvin + \myvshock$   & $\myxshock$   &  $\langle n \rangle$  \\
                              & {\small (km s$^{-1}$)} & {\small (pc)} &  {\small (cm$^{-3}$)} \\
 \hline 
 \hline
3D with cooling (C2D1000)     & 21.276                 & 6.183         & 6.750  \\
1D with $\mygamma$ (E2D0000)  & 22.067                 & 6.338         & 6.081  \\

            \hline
            \hline
        \end{tabular}
    }\par
    \bigskip
            \textbf{Note.} $\myvin+\myvshock$, $\myxshock$,
            and  $\langle n \rangle$ measured in
            Runs C2D1000 and E2D0000.
    \label{table:geff}
\end{table}

Note that the implementation of this effective EoS is unfortunately not that straightforward.
For example,
to introduce a time-evolving $\mygamma$, 
we would like to 
measure shock propagation speed and elapsed-time 
since the last shock passage even in large-scale simulations.
This is, however, computationally expensive
and time-consuming, similar to following stellar population evolution below the spatial resolution
to blow supernovae at a correct timing.
Such difficulties have to be also discussed and left for future studies.
Nevertheless, since the typical frequency of shock passages is as high as once per Myr 
\citep[\eg,][; see also Section~\ref{subsec:diffset}]{McKee1977},
we may expect that the quasi-steady $\mygamma$ at $\sim 3$ Myr is still close to the typical time-averaged state of the actual ISM\@.

\subsection{Limitations of Current Converging Flow Systems}
\label{subsec:diffset}
In this section, we briefly address some potential limitations/issues
in converging flow simulations (not only ours but also in general).
Given that 1 Myr is the typical interval in the ISM between 
successive shock passages by multiple supernovae and/or 
H {\sc ii} regions \citep{McKee1977,Inutsuka2015},
supersonic flow in reality continue in a fixed direction only $\leq$ 1 Myr.
Statistically speaking, successive flows essentially propagate from any direction
and they
incident the shock-compressed layer at some angle.
Almost all of the converging flow studies therefore
presumably keep the flow injection too long in a fixed direction.
There are previous studies
introducing an inclination angle between flows 
to investigate the effect of magnetic diffusion and supercritical core formation \citep{Kortgen2015}
and to investigate the reorientation of pre-existing filaments
\citep[\cf,][]{Fogerty2016,Fogerty2017},
but it is still left for future studies
to reveal how the mean properties of the shock-compressed layer depend on
such inclinations and multiple compressions
by flows from various angles.

Similarly, the typical dynamical timescale of the shock-compressed layer is a few Myr 
(\eg, the crossing time of the WNM component over the shock-compressed layer is $10$ pc / $10 \kms$
in Run C2D1000).
It is thus also left for future studies 
to investigate how an already-created shock-compressed layer
expands and/or shrinks once the inflow ceases and to measure whether 
the turbulence and $f_{\rm CNM}$ is maintained or not.
Limiting the inflow mass is one of the technique to study such condition;
\cite{vazquezsemadeni2007} for example demonstrate that 
the global and local collapses of the shock-compressed layers occur
once all the gas finish accreting onto the shock-compressed layers.

The concept of converging flow setup is
to easily perform calculations in the post-shock rest frame.
However, most of the post-shock regions in reality is presumably 
not sandwiched by two shock fronts as in the converging flows,
but instead by one shock front and one contact discontinuity.
Converging flow is just an analogue 
of such a shock-contact discontinuity system,
and only one half side of shock-compressed layer is meaningful.
The two-shock-front configuration likely 
impacts
the time-evolution of the shock-compressed layer.
For example as shown in Figure~\ref{fig:f24f25f26f27},
fast WNM flows continue deep into the layer
and interact each other, especially when $\mydrhop$ is large,
which depends on the shock front geometry on both two sides.
The turbulent properties and $f_{\rm CNM}$ may accordingly differ
in a 
shock-contact discontinuity system.
Simulation studies of a shock-contact discontinuity system
\cite[\eg,][]{Koyama2002} is still limited 
and careful comparison is left for future studies.

We also ignore magnetic fields for simplicity in this article, 
but they play a pivotal role; for example 
magnetic field pressure supports the shock-compressed
layer and the turbulence decays, especially in case
the field lines have perpendicular orientation against 
the inflow \citep{Heitsch2009,vazquezsemadeni2011,Inoue2012,Iwasaki2018}.
We expect that magnetic fileds do not modify our proposed $\mygamma$ significantly
because the magnetized shock-compressed layer tends to be equipartition 
\citep[see][for $\myvin < 20 \, \kms$ cases]{Iwasaki2018},
but this still has to be investigated with magnetized converging flow simulations.

\section{Summary}
\label{sec:concl}
We perform a series of hydrodynamics simulations of converging warm neutral medium (WNM) flows
with heating and cooling,
to calculate the cold neutral medium (CNM) formation and
to investigate the mean physical properties of the multiphase interstellar medium (ISM) 
averaged over the shock-compressed layer on a 10 pc scale,
such as the mean shock front position $\myxshock$, the mean density $\langle n \rangle$,
and the density-weighted velocity dispersion $\sqrt{\langle \delta v_{\rm dw}^2\rangle_{\rm tot}}$.
Under a fixed flow velocity of $20 \kms$ and the Kolmogorov power spectrum 
in the upstream density fluctuation,
we systematically vary the amplitude of the 
upstream density 
fluctuation $\mydrhop = \sqrt{\langle \delta \rho_0 \rangle}/\rho_0$,
random phases of the fluctuation $\alpha$, and the spatial resolution $\Delta x$.
We find that two distinct post-shock states exist depending on $\mydrhop$,
typically divided by $\mydrhop=10$ \%.
We list our main findings as follows.

\begin{enumerate}
    \item The convergence in $\myxshock$ and $\langle n \rangle$ requires $\Delta x =0.02$ pc 
              by fully resolving the typical cooling length on which the phase transition occurs from the WNM to CNM.
    \item The trend of convergence, however, differs depending on $\mydrhop$. 
              The trend is non-monotonic when $\mydrhop > 10$ \% due to the intrinsic large variation
              induced by different phase $\alpha$ and different $\Delta x$, and 
              calculations with coarse resolutions of $\Delta x > 0.02$ pc practically provide simliar values in $\myxshock$ and $\langle n \rangle$.
              The convergence in the case of $\mydrhop \leq 10$ \% is monotonic and stringently requires $\Delta x =0.02$ pc.
    \item The significant deformation of the shock fronts in large $\mydrhop$ cases 
              drive strong turbulence up to $\sqrt{\langle \delta v_{\rm dw}^2\rangle_{\rm tot}} \sim 7\,\kms$, 
              which prevents the dynamical condensation by cooling and the CNM formation.
              When $\mydrhop$ is small, the shock fronts maintain a straight geometry
              and the velocity dispersion is limited to the thermal-instability mediated level 
              of $\sqrt{\langle \delta v_{\rm dw}^2\rangle_{\rm tot}} = 2$ -- $3\,\kms$.
    \item The shock-compressed layer is wider (narrower) and less dense (denser) with larger (smaller) $\mydrhop$,
              where the CNM mass fraction is $\sim 45$ \% and $\sim 70$ \%
              when $\mydrhop = 31.6$ \% and $3.16$ \%, respectively.
              %the WNM:CNM mass ratio
              %is $\sim 4:6$ and $\sim 1:9$ 
    \item The turbulent energy supports the shock-compressed layer when $\mydrhop > 10$ \%, whereas
              both the turbulent and thermal energy equally supports the shock-compressed layer when $\mydrhop \leq 10$ \%.
    \item We formulate an effective equation of state, $P \propto \rho^{\mygamma}$,
              which approximates the multiphase ISM as a one-phase medium.
              $\mygamma$ measured from our converging-flow simulations
              ranges from $0.9$ (with large $\mydrhop$) to $0.7$ (with small $\mydrhop$), softer than isothermal.
\end{enumerate}
These results have to be further investigated by simulating
other shock orientations, ceasing of mass accretion, 
and by including magnetic fields as well as in a shock-contact discontinuity system.
We also hope that 
upcoming observations (\eg, ALMA, SKA, ngVLA)
constrain the formation condition of the multiphase ISM,
such as $\mydrhop$, by measuring
the physical properties of the ISM (velocity dispersion, the CNM mass fraction, \etc).

\section*{ACKNOWLEDGMENTS}
We are grateful to the anonymous reviewer for his/her careful reading and comments,
which improved our manuscript significantly.
Numerical computations were carried out on Cray XC30 and XC50 
at Center for Computational Astrophysics, National Astronomical Observatory of Japan.
MINK (15J04974, 18J00508, 20H04739), TI (18H05436, 20H01944),
SI (16H02160, 18H05436, 18H05437), KT (16H05998, 16K13786,17KK0091),
KI (19K03929), and KEIT (19H05080, 19K14760)
are supported by Grants-in-Aid from the Ministry of Education, Culture,
Sports, Science, and Technology of Japan. 
KT and KEIT are also supported by NAOJ ALMA Scientific Research grant No. 2017-05A.
%MINK appreciate Atsushi J. Nishizawa and Chiaki Hikage for helping our data analysis.
MINK is grateful to Eve C. Ostriker, Woong-Tae Kim,
Patrick Hennebelle, Philippe Andr{\'e}, 
Marc-Antoine Miville-Desch{\^e}nes, Antoine Marchal,
Jin Koda, Kentaro Nagamine, Tomoyuki Hanawa, Kazuyuki Omukai, and Shinsuke Takasao
for fruitful comments.
%MINK also appreciate Tomoyuki Hanawa, Kohji Tomisaka, Tetsuo Hasegawa, Kazuyuki Omukai, 
%Eve C. Ostriker, Woong-Tae Kim,
%Patrick Hennebelle, Alvaro Hacar, Juan Soler, 
%Jin Koda, Marc-Antoine Miville-Deschenes, Valeska Valdivia, Antoine Marchal,
%Akio K. Inoue, Kentaro Nagamine, Tomohiro Ono, 
%Shinsuke Takasao,
%Naoshi Sugiyama, Kiyotomo Ichiki, Misato Fukagawa, and Kengo Tachihara for fruitful comments.

 %-------------- BIBLIO -------------------------------------------------------

%\bibliographystyle{apj}
\bibliographystyle{aasjournal}
\bibliography{galaxy_new}

% Zel'dovich, Ya. B., & Pickel'ner, S.B.
% "The Phase Equilibrium and Dynamics of a Gas Volume that is Heated and Cooled.
% http://www.jetp.ac.ru/cgi-bin/e/index/e/29/1/p170?a=list

 %-----------------------------------------------------------------------------

\end{document}